\documentclass[journal,twoside,web]{ieeecolor}

\usepackage{generic}
\usepackage{graphicx}
\usepackage[outdir=./]{epstopdf}
\usepackage{amsmath,amssymb,amsfonts}
\usepackage[breaklinks=true,colorlinks,bookmarks=false]{hyperref}
\usepackage{textcomp}
\usepackage{multirow}
\usepackage{subcaption}
\usepackage[linesnumbered,ruled,vlined]{algorithm2e}

\usepackage[margin=1in]{geometry}

\graphicspath{{Figures/}}

\begin{document}
\title{Spatial-aware Transformer-GRU Framework for Enhanced Glaucoma Diagnosis from 3D OCT Imaging}
\author{Mona~Ashtari-Majlan, David~Masip,~\IEEEmembership{Senior~Member,~IEEE}
\thanks{This work was supported by the Spanish Ministry of Science and Innovation through the PID2022-138721NB-I00 grant, funded by the FEDER initiative.}
\thanks{M. A. and D. M. are with the Department of Computer Science, Multimedia, and Telecommunications, Universitat Oberta de Catalunya, Barcelona, Spain (e-mail: \{mashtarimajlan,~dmasipr\}@uoc.edu)}
}

\maketitle

\begin{abstract}
Glaucoma, a leading cause of irreversible blindness, necessitates early detection for accurate and timely intervention to prevent irreversible vision loss. In this study, we present a novel deep learning framework that leverages the diagnostic value of 3D Optical Coherence Tomography (OCT) imaging for automated glaucoma detection. In this framework, we integrate a pre-trained Vision Transformer on retinal data for rich slice-wise feature extraction and a bidirectional Gated Recurrent Unit for capturing inter-slice spatial dependencies. This dual-component approach enables comprehensive analysis of local nuances and global structural integrity, crucial for accurate glaucoma diagnosis. Experimental results on a large dataset demonstrate the superior performance of the proposed method over state-of-the-art ones, achieving an F1-score of 93.01\%, Matthews Correlation Coefficient (MCC) of 69.33\%, and AUC of 94.20\%. The framework's ability to leverage the valuable information in 3D OCT data holds significant potential for enhancing clinical decision support systems and improving patient outcomes in glaucoma management.
\end{abstract}


\section{Introduction}
\label{sec:introduction}
\IEEEPARstart{G}{laucoma}, a leading cause of irreversible blindness worldwide~\cite{steinmetz2021causes}, is particularly insidious due to its asymptomatic nature in the early stages. This characteristic underscores the critical importance of timely diagnosis, as early intervention is vital for preventing visual impairment and irreversible vision loss~\cite{Weinreb2014Review}. Optical Coherence Tomography (OCT), a three-dimensional (3D) non-invasive imaging modality, has revolutionized ophthalmological diagnostics by enabling high-resolution visualization of the eye's intricate anatomy~\cite{huang1991optical}. This technology has proven valuable for the early detection and management of glaucoma.

Despite the widespread adoption of OCT imaging, a significant proportion of studies on glaucoma diagnosis have primarily focused on analyzing two-dimensional (2D) OCT scans~\cite{Gabriel2020Glaucoma, garcia2021circumpapillary}. These approaches typically concentrate on the middle slice of the OCT volume, centered around the Optic Nerve Head (ONH), which has been demonstrated to be effective in characterizing structural changes related to glaucoma~\cite{WANG2020101695}. However, this localized analysis inherently overlooks the potential diagnostic value contained within the comprehensive three-dimensional data provided by OCT imaging.

Glaucoma is a multifaceted disease that can manifest progressive structural alterations throughout the retina, including notable thinning of the Retinal Nerve Fiber Layer (RNFL)~\cite{JONAS20172183}. This widespread impact suggests that analyzing the total B-scan slices within the OCT volume could reveal important characteristics and patterns indicative of glaucoma. Consequently, a holistic analysis that leverages the rich information within the entire ocular structure is crucial for enhancing diagnostic accuracy and enabling early intervention\cite{hemelings2021deep, Maetschke2019feature, RAN2019e172}.

Herein lies the potential of Artificial Intelligence (AI)-based clinical decision support systems, which can assist practitioners by leveraging advanced algorithms for automated disease detection and management. To address the challenges associated with the manual inspection of volumetric OCT data, in this work, we propose a novel deep learning-based framework designed to maximize the diagnostic potential of 3D OCT imaging for automated glaucoma screening. In contrast to traditional analysis techniques that often overlook valuable information, our approach systematically extracts and integrates features from the entire OCT scan, revealing subtle glaucomatous indicators distributed across the ocular structure.

\begin{figure*}[!htbp]
  \centering
  \includegraphics[width=0.95\linewidth]{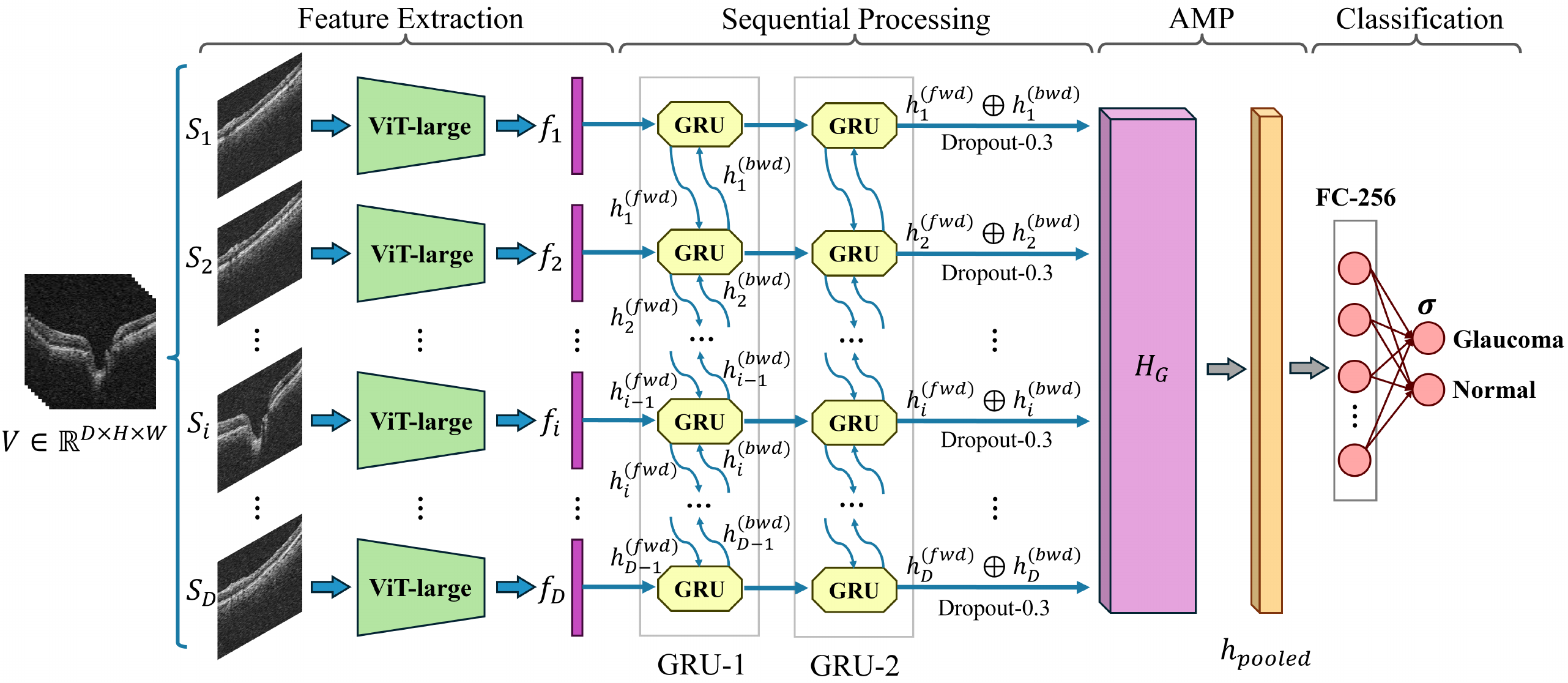}
  \caption{Schematic of the proposed framework for detecting glaucoma. The framework uses a large Vision Transformer (ViT-large) feature extractor called RETFound, which was pre-trained on a large OCT dataset. The symbol $\oplus$ represents the concatenation of spatial states at timestep $i$, combining information from anterior and posterior spatial locations to capture long-range spatial dependencies. The framework includes a Bidirectional Gated Recurrent Unit (GRU), Adaptive Max Pooling (AMP), and a Fully Connected (FC) layer.}
  \label{fig:Schematic}
\end{figure*}

The proposed end-to-end framework comprises two core components: (1) a Transformer-based model for extracting features from individual slices within the OCT volumes and (2) a Recurrent Neural Network (RNN) for integrating these features into a comprehensive representation of the retinal structure. This dual-component architecture (depicted in Fig.~\ref{fig:Schematic}) enables the model to capture both local nuances present in individual slices and the global structural integrity of the retina while incorporating inter-slice dependencies essential for a comprehensive analysis. 

%%%%%%%%%PLACEMENT

We used a pre-trained large Vision Transformer (ViT-large) backbone developed by Zhou et al.~\cite{zhou2023foundation} for feature extraction from individual B-scan slices. This foundational model, called RETFound, leverages self-supervised learning on 1.6 million unlabeled retinal images to acquire generalizable representations. Notably, this model exhibits proficiency in identifying intricate patterns and features relevant to ocular health, ensuring high-precision analysis of individual slices within 3D OCT images. After feature extraction, the RNN component, implemented using bidirectional Gated Recurrent Units (GRUs), integrates these slice-based representations to account for the sequential and interconnected nature of slices within the three-dimensional space. This integration is crucial for capturing the dynamic spatial relationships and dependencies between slices, thereby reflecting the comprehensive structural complexity of the retina. Experimental results on a 3D OCT dataset~\cite{Maetschke2019feature} validate the superior performance of our proposed framework in glaucoma detection compared to state-of-the-art methods.

The following is the organization of the paper: Section~\ref{sec:literature} reviews the previous studies. Section~\ref{sec:proposed} describes the proposed method. In Section~\ref{sec:experiment}, we present and discuss the experimental setup and results. Finally, in Section~\ref{sec:conclusion}, we provide the conclusion, summarizing our findings and suggesting future research directions.

\section{Literature review}
\label{sec:literature}
Glaucoma is a medical condition characterized by structural changes in the optic nerve head and thinning of the retinal nerve fiber layer in the retina~\cite{NouriMahdavi2021}. Numerous studies have demonstrated the capability of OCT imaging to detect these structural alterations, facilitating early diagnosis and monitoring of the disease~\cite{chen2019combination, JUNEJA2022117202, lee2020diagnosing, MEDEIROS2019513, Raja2021Clinically}. A comprehensive literature review covering the use of OCT imaging for glaucoma diagnosis is available in~\cite{ashtarimajlan2023deep}.

Numerous previous studies have focused on examining two-dimensional (2D) OCT scans, also known as B-scans, for glaucoma detection~\cite{Gabriel2020Glaucoma}. These studies have explored various deep learning methods, integrating ophthalmological domain knowledge to enhance the diagnostic efficacy of OCT imaging. For instance, Raja et al.~\cite{Raja2021Clinically} developed a deep convolutional autoencoder network to segment RNFL regions in the retina. The region-aware encoder component of their proposed model was employed for glaucoma detection, achieving an accuracy of 94.91\%, F1-score of 95.77\%, and an area under the curve (AUC) of 98.71\% on the AFIO dataset~\cite{RAJA2020105342}. Additionally, Garcia et al.~\cite{garcia2021circumpapillary} developed a hybrid neural network incorporating a VGG16 backbone tailored for processing raw OCT B-scans and an RNFL descriptor for extracting thickness information in this region. Their approach achieved an accuracy of 87.88\% and an F1-score of 81.82\% for discriminating between healthy, early, and advanced glaucoma samples. Lee et al.~\cite{lee2020diagnosing} also used the features extracted from maps representing SD-OCT Ganglion Cell Inner Plexiform Layer (GCIPL) and RNFL thicknesses and deviations for glaucoma diagnosis. They achieved an AUC of 99\%, a sensitivity of 94.70\%, and a specificity of 100\% when tested on a private dataset comprising 350 glaucomatous and 307 healthy SD-OCT image sets.

The availability of pre-trained deep learning models has facilitated the transfer of features learned from large-scale 2D datasets, such as ImageNet, for more accurate classification of 2D OCT scans. However, one of the primary limitations of relying solely on 2D OCT scans for glaucoma detection is that a single 2D slice can only capture a static view of the retina, failing to account for the complex nature of RNFL thinning and damage distribution across the retinal structure~\cite{WANG2020101695}. In an attempt to address this limitation, Wang et al.~\cite{WANG2020101695} proposed selecting every three successive B-scan images from a 3D OCT volume to form a three-channel input with volume-level labeling. They curated the Hong Kong dataset, comprising 2,926 glaucomatous and 1,951 normal OCT scans, and employed it to train their model. This approach achieved an accuracy of 92.70\%, an F1-score of 94.10\%, and an AUC of 97.70\% for diagnosing glaucoma. However, their methodology did not fully consider inter-slice dependencies, a critical aspect of spatial context that could potentially improve the accuracy of glaucoma detection.

To better capture the intricacies of 3D OCT scans, several studies have employed three-dimensional convolutional neural network (3D-CNN) models~\cite{George2020IEEE, Maetschke2019feature, Noury2022Deep, Russakoff2020Deep}. For instance, Noury et al.~\cite{Noury2022Deep} proposed a 3D-CNN model with multiple layers of dense and 3D spatial convolutional blocks. They curated the Stanford dataset, comprising 1,617 glaucomatous and 844 normal OCT scans, for training their model. Subsequently, they evaluated their proposed model on this dataset and three additional external datasets, achieving an AUC of 91.00\% for the Stanford dataset, 80.00\% for the Hong Kong dataset, 94.00\% for the India dataset, and 87.00\% for the Nepal dataset. Furthermore, Ran et al.~\cite{RAN2019e172} proposed training a 3D ResNet34 model from scratch on the Hong Kong dataset for glaucoma diagnosis, yielding an AUC of 96.90\% and accuracy of 91.00\% on the primary validation set. George et al.~\cite{George2020IEEE} proposed a novel model for glaucoma diagnosis and VF Index estimation. The model is an attention-guided 3D-CNN with three streams, where one stream receives raw 3D-OCT scans as inputs, while the inputs for the other two streams are determined during training, guided by attention heatmaps from 3D Grad-CAM. The model was trained on a dataset containing 3,355 glaucomatous and 427 normal OCT scans. The results showed that the model achieved an AUC of 93.77\%, an accuracy of 91.07\%, and an F1-score of 94.88\% for glaucoma diagnosis.

Training 3D-CNN architectures for analyzing 3D images is computationally intensive, posing efficiency challenges and the risk of the curse of dimensionality. To mitigate these issues, some studies in medical imaging have proposed decomposing 3D images into a sequence of 2D slices~\cite{GARCIA2021105855, He2020Brain, Kurmann2019Fused, Liu2018classification, Rashid2023Biceph, VANKDOTHU2022107960}. This strategy aims to balance computational burden and depth of analysis while preserving the crucial inter-slice context for accurate image classification. Specifically for diagnosing glaucoma, Garc{\'\i}a et al.~\cite{GARCIA2021105855} proposed a method that combines CNNs and RNNs with Long Short-Term Memory (LSTM), using a pre-trained VGG16 model to extract discriminative features from 2D slices of OCT images. They achieved an accuracy of 81.25\%, an F1-score of 78.57\%, and an AUC of 80.79\% on a dataset containing 144 glaucomatous and 176 normal OCT scans.

Other data modalities have also been used to further improve the accuracy of glaucoma diagnosis. For instance, Song et al.~\cite{Song2021Deep} proposed a transformer-based architecture that uses OCT and visual field (VF) results to diagnose glaucoma. Their model incorporated deep reasoning to examine implicit pairwise relations between these two complementary modalities. The experimental results, conducted on a dataset comprising 697 glaucomatous and 698 normal scans, demonstrated an accuracy of 88.30\%, an F1-score of 88.90\%, and an AUC of 93.90\% for glaucoma diagnosis.

Our approach deviates from previous studies in several key aspects, offering enhanced diagnostic capabilities. Specifically, we employ a transformer-based feature extractor trained on a large-scale OCT dataset, surpassing the performance of models pre-trained on ImageNet and fine-tuned on relatively small OCT datasets. Additionally, using a bidirectional GRU model for analyzing OCT slices ensures a more comprehensive capture of anterior and posterior spatial dependencies between slices. Furthermore, our model is trained on a larger yet imbalanced dataset, better reflecting real-world clinical scenarios. These distinctions underscore our model's superior generalizability and applicability in accurately diagnosing glaucoma through OCT image analysis.

\section{Proposed Method}
\label{sec:proposed}

The proposed model is designed for end-to-end glaucoma classification. It integrates two core components: Feature Extraction and Sequential Processing. The model employs a visual transformer and an RNN with the gated recurrent unit to enhance the efficiency and accuracy of glaucoma diagnosis. By leveraging these techniques, the model aims to harness the diagnostic potential of 3D OCT scans for reliable glaucoma detection.

\subsection{Pre-processing}
We performed a two-step pre-processing to standardize the input data for optimal integration with our proposed model.
Initially, the scans underwent normalization using the mean and standard deviation of ImageNet. This step is crucial for adjusting pixel intensity values to a common scale, thereby enhancing the model's ability to learn meaningful patterns from the data. Subsequently, the scans were resized to a uniform dimension of 64$\times$128$\times$128, ensuring consistency across the dataset and aligning with the input specifications of the pre-trained components of our model.

\subsection{Feature Extraction}

The Feature Extraction component uses the ViT-large encoder from the RETFound model~\cite{zhou2023foundation}, a self-supervised autoencoder framework pre-trained on large-scale OCT data. The RETFound encoder employs a ViT-large architecture~\cite{dosovitskiy2021an} with 24 Transformer blocks and a 1024-dimensional embedding vector. We discard the decoder layers for our 3D OCT scans and leverage the pre-trained ViT-large encoder to extract rich feature representations.

Mathematically, let $V$ denote a 3D OCT volume, which is decomposed into $D$ individual slices, where the $i$-th slice is represented as $S_i \in \mathbb{R}^{H \times W}$, for $i = 1, 2, \dots, D$. Each slice $S_i$ is independently processed by the ViT-large model to derive a feature vector $f_i \in \mathbb{R}^{1024}$, thereby extracting rich, slice-specific feature representations from the 3D OCT data.

\subsection{Sequential Processing}

Following the feature extraction, to encompass all inter-slice correlations and spatial dependencies intrinsic to OCT imaging, we propose using a bidirectional GRU architecture for glaucoma classification. The GRU, a variant of RNNs tailored for sequence modeling, offers an effective framework for analyzing sequential data. In our methodology, each feature vector $f_i$ extracted from slice $S_i$ of the OCT scan constitutes a distinct state at timestep $i$ within the sequence. This bidirectional approach enables the GRU layers to adeptly capture both anterior and posterior spatial dependencies among the sequential OCT slices.
Such a comprehensive analysis facilitates an in-depth understanding of the spatial relationships and pathological indicators critical for accurate glaucoma diagnosis. Moreover, the GRU architecture efficiently mitigates the vanishing gradient problem and enables the learning of long-term dependencies with fewer parameters than conventional long-term memory models, thereby enhancing the model's efficiency and effectiveness in glaucoma classification.

To further delineate the sequential processing component of our proposed framework, the architecture employs two consecutive bidirectional GRU layers. Given a sequence of feature vectors $\{f_1, f_2, \cdots, f_D\}$ where $D = 64$, the bidirectional GRU processes each $f_i$ using Eq.~\ref{eq:01}.

\begin{equation}
    \label{eq:01}
    \begin{aligned}
    h_i^{\mathrm{(fwd)}} &= \mathrm{GRU}^{\mathrm{(fwd)}}(f_i, h_{i-1}^{\mathrm{(fwd)}}) \\
    h_i^{\mathrm{(bwd)}} &= \mathrm{GRU}^{\mathrm{(bwd)}}(f_i, h_{i+1}^{\mathrm{(bwd)}})
    \end{aligned}
\end{equation}

where $h_i^{\mathrm{(fwd)}}$ and $h_i^{\mathrm{(bwd)}}$ denote the forward and backward hidden states at timestep $i$, respectively. The final hidden states are concatenated to form a comprehensive sequence representation, which is expressed by Eq.~\ref{eq:M2}.

\begin{equation}
    \label{eq:M2}
    \resizebox{.9\columnwidth}{!}{$
      H_G = \left[h_1^{\mathrm{(fwd)}} \oplus h_1^{\mathrm{(bwd)}}, h_2^{\mathrm{(fwd)}} \oplus h_2^{\mathrm{(bwd)}}, \ldots, h_D^{\mathrm{(fwd)}} \oplus h_D^{\mathrm{(bwd)}}\right]
  $}
\end{equation}
where $\oplus$ denotes the concatenation operation. After a Dropout layer, this sequence $H_G$ is then aggregated along the spatial dimension to form a unified feature representation $h_{\mathrm{pooled}}$ through an Adaptive Max Pooling (AMP) layer. This layer effectively extracts the essential features from the sequence, capturing the most important information across all timesteps. Finally, a Fully Connected (FC) layer maps the aggregated feature representation to the probability distribution over the two classes, i.e., glaucoma and normal, represented in Eq.~\ref{eq:M3}.

\begin{equation}
    \label{eq:M3}
    p = \sigma\left(W' \cdot h_{\mathrm{pooled}} + b'\right)
\end{equation}
where $p$ denotes the probability vector associated with the classification of a specific volume $V$, $\sigma$ denotes the Sigmoid activation function, $W'$ and $b'$ are the weight and bias of the FC layer, respectively.

We used the Focal Loss function to train our proposed model, which is a variant of the binary cross entropy loss that addresses the class imbalance issue by adjusting the loss contribution from each sample~\cite{Lin_2017_ICCV}. This method focuses more on hard-to-classify examples, ultimately reducing the impact of class imbalance on the model's learning process. The Focal Loss is defined in Eq.~\ref{eq:M4}.

\begin{equation}
    \label{eq:M4}
    \mathrm{FL}(p_t) = -\alpha_t (1 - p_t)^\gamma \log(p_t)
\end{equation}
where $p_t$ is the model's estimated probability for the class with label $t$, $\alpha_t$ is a weighting factor to counteract class imbalance, and $\gamma$ controls the rate at which easy examples are down-weighted. Note that the details for identifying our model's optimal set of hyper-parameters are presented in Section~\ref{sec:parametertuning}. The proposed glaucoma classification framework is outlined in Algorithm~\ref{alg:proposed}, and the implementation code is available at https://github.com/Mona-Ashtari/SpatialOCT-Glaucoma.

\begin{algorithm}[!htbp]
\caption{Proposed Glaucoma Classification framework}
\label{alg:proposed}
\KwIn{3D OCT volume $V \in \mathbb{R}^{D \times H \times W}$}
\KwOut{Glaucoma diagnosis Probability $p \in [0, 1]$}

\textbf{Feature Extraction:}\\
Decompose $V$ into $D$ slices: $\{S_1, S_2, \dots, S_D\}$.\\
\For{$i = 1$ to $D$}{
    $f_i \gets \text{Pre-trained ViT-large}(S_i)$ \tcp{Extract feature vector $f_i \in \mathbb{R}^{1024}$.}
}

\textbf{Sequential Processing:}\\

Pass the sequence $\{f_1, f_2, \dots, f_D\}$ through two stacked bidirectional GRU layers:\\
$H_G \gets \text{StackedGRUs}(\{f_1, f_2, \dots, f_D\})$ \tcp{Sequence of GRU outputs.}

\textbf{Feature Pooling:}\\
$h_{\mathrm{pooled}} \gets \text{AdaptiveMaxPooling}(H_G)$\\

\textbf{Classification:}\\
$p \gets \sigma(W' \cdot h_{\mathrm{pooled}} + b')$ \tcp{Sigmoid activation.} %\\

\textbf{Loss Calculation:}\\
$\mathrm{FL}(p) = -\alpha \cdot (1 - p)^\gamma \cdot \log(p)$ \tcp{Focal Loss.} %\\

\textbf{Model Training:}\\
Optimize model parameters using the Adam optimizer with the Focal Loss.
\end{algorithm}

\section{Experiments}
\label{sec:experiment}

\subsection{Dataset}

In this study, we used a dataset introduced by Maetschke et al.~\cite{Maetschke2019feature}\footnote{https://doi.org/10.5281/zenodo.1481223}, comprising 3D OCT scans centered on the optic nerve head. The dataset includes scans from 624 patients obtained using a Cirrus SD-OCT Scanner (Zeiss, Dublin, CA, USA), with physical dimensions of 6$\times$6$\times$2 mm and a resolution of 
64$\times$64$\times$128
% 200$\times$200$\times$1024 
voxels per volume. Only scans with a signal strength of 7 or higher were included, resulting in a total of 1110 scans. These scans maintain their original laterality, avoiding any left to right eye alterations. Among these, 263 scans were classified as healthy, and 847 were diagnosed with primary open-angle glaucoma, based on glaucomatous visual field defects confirmed by at least two consecutive abnormal test results. Figure~\ref{fig:glaucoma_vs_healthy} illustrates the first, middle, and last slices of 3D OCT samples from both healthy and glaucoma cases. The demographic details like gender and race distribution, along with mean values and standard deviations for patient's age, Intraocular Pressure (IOP), Mean Field Defects (MD), and Glaucoma Hemifield Test (GHT)~\cite{Peter1992Hemifield} results are presented in Table~\ref{tab:data_demo}. It should be noted that demographic data was incomplete for some patients, and as a result, the aggregate numbers may not align exactly with the overall dataset size.
For a detailed description of the dataset, you can refer to the Maestack et al~\cite{Maetschke2019feature}.

\begin{table}[!htbp]
\centering
\caption{Demographic details of the dataset}
\label{tab:data_demo}
\begin{tabular}{lcc}
\hline
       & Healthy                     & Glaucoma                    \\ \hline
Female & 88                          & 217                         \\
Male   & 49                          & 215                         \\
White  & 101                         & 318                         \\
Black  & 30                          & 154                         \\
Asian  & 5                           & 12                          \\
Age    & 54.1 ± 15.3 {[}22.1-88.9{]} & 64.3 ± 12.5 {[}25.2-93.8{]} \\
IOP    & 13.5 ± 2.4 {[}9-23{]}       & 16.7 ± 5.8 {[}2-51{]}       \\
MD     & -0.8 ± 1.7 {[}-9.9-2.8{]}   & -6.8 ± 8.1 {[}-32.9-2.17{]} \\
GHT    & 1.6 ± 1.0 {[}1-6{]}         & 2.4 ± 0.9 {[}1-6{]}         \\ \hline
\end{tabular}
\end{table}

In our study, we employed a 5-fold cross-validation approach to ensure our model's robustness and generalizability across different subsets of the data. The rationale behind using 5-fold cross-validation is to provide a more reliable estimate of the model's performance on unseen data by averaging results across multiple folds. The average number of samples allocated for training, validation, and testing in each fold is detailed in Table~\ref{tab:dataset}. To ensure that the scans from the same patient are included in the same subset, the data split in each fold is performed at the subject level.

\begin{table}[!htbp]
\centering
\caption{Dataset summary}
\label{tab:dataset}
\begin{tabular}{lcccc}
\hline
\multirow{2}{*}{} & \multirow{2}{*}{Total} & \multicolumn{3}{c}{5-fold cross validation} \\ \cline{3-5} 
                  &                        & Train       & Validation       & Test       \\ \hline
Glaucoma          & 847                    & 609.2         & 68.4               & 169.4        \\
Normal            & 263                    & 189.2         & 21.2               & 52.6         \\
Total             & 1110                   & 798.4         & 89.6               & 222        \\ \hline
\end{tabular}
\end{table}

\subsection{Implementation details and Evaluation metrics}
\label{sec:parametertuning}

The proposed framework is implemented using Python based on the PyTorch library version 1.8.1 on a computer powered by Intel(R) Xeon(R) Silver 4210R CPU with 16G of RAM and the NVIDIA GeForce RTX 2080 GPU with 12G of RAM.
We trained the proposed framework using Adam optimizer~\cite{kingma2014adam}, configured with the first momentum of 0.9, the second momentum of 0.999, and the stability constant of $10^{-8}$. 
To ensure efficient convergence, we explored the impact of varying combinations of learning rates and batch sizes on model performance across the $\{10^{-5}, 5\times10^{-5}, 10^{-4}, 5\times10^{-4}, 10^{-3}\}$ set for learning rates and $\{8, 16, 32, 64, 128\}$ set for batch sizes. 
The optimal performance on the validation set was achieved with an initial learning rate of $10^{-4}$ and a batch size of 16. We employed a learning rate scheduler to adjust the learning rate by a factor of 0.9 every 5 epochs. 
To address the challenge of data imbalance in our dataset, we adopted a balanced batch training strategy to ensure fair class representation and mitigate bias.
The model was trained for a maximum of 100 epochs and incorporated early stopping triggered by validation loss stagnation after 6 epochs, which improved computational efficiency and prevented overfitting.
We conducted an empirical exploration to find the optimal combination of five other hyper-parameters for improving the performance of the proposed glaucoma classification framework. The hyper-parameters include the hidden size of two GRU layers, the model dropout rate, and the  $\alpha$ and $\gamma$ parameters in the Focal Loss function.

One important parameter for determining the capacity of the GRU architecture to learn and represent data is the hidden size of the GRU layers. Given that the architecture includes two GRU layers, we evaluated different configurations for the hidden sizes, ranging from 512, 256, to 128 on the validation set. The summary of the results is presented in Table~\ref{tab:GRUsize}. Notably, as the size of the hidden layers decreases from 512 to 128, there is a general trend toward reduced F1-scores, indicating that larger hidden layer sizes tend to capture more complex patterns and dependencies in the data. However, the optimal configuration is achieved with the first GRU layer of size 256 and the second of size 128, resulting in the highest F1-score of 93.12\%. This suggests a threshold for increasing layer size, highlighting the importance of balancing model capacity and the risk of overfitting.

\begin{table}[!htbp]
\centering
\caption{F1-score (\%) with respect to the hidden sizes of GRU layers}
\label{tab:GRUsize}
\begin{tabular}{lcccc}
\hline
\multicolumn{1}{l}{\multirow{2}{*}{\textbf{GRU-1}}} & \multicolumn{4}{c}{\textbf{GRU-2}} \\ \cline{2-5} 
 & \textbf{512}     & \textbf{256}    & \textbf{128}    & \\ \hline
\textbf{512} & 92.33   & 92.20  & 91.36 & \\
\textbf{256} & -       & 91.03  & \textbf{93.12} & \\
\textbf{128} & -       & -      & 92.05 & \\ \hline
\end{tabular}
\end{table}

\begin{figure*}[!htbp]
  \centering
  \begin{subfigure}{0.4\linewidth}
    \includegraphics[width=\linewidth]{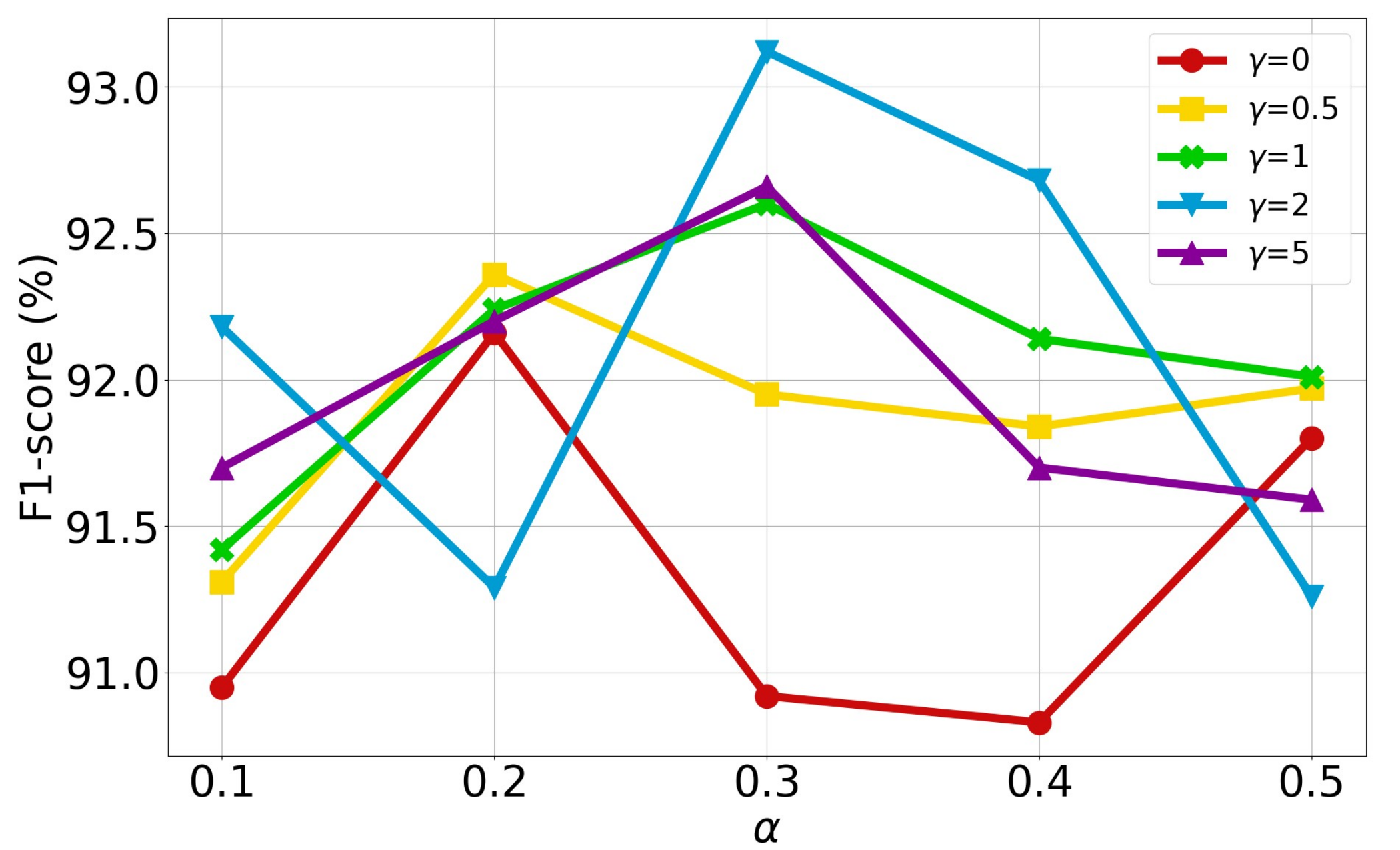}
    \caption{}
    \label{fig:subalpha}
  \end{subfigure}
  \begin{subfigure}{0.4\linewidth}
    \includegraphics[width=\linewidth]{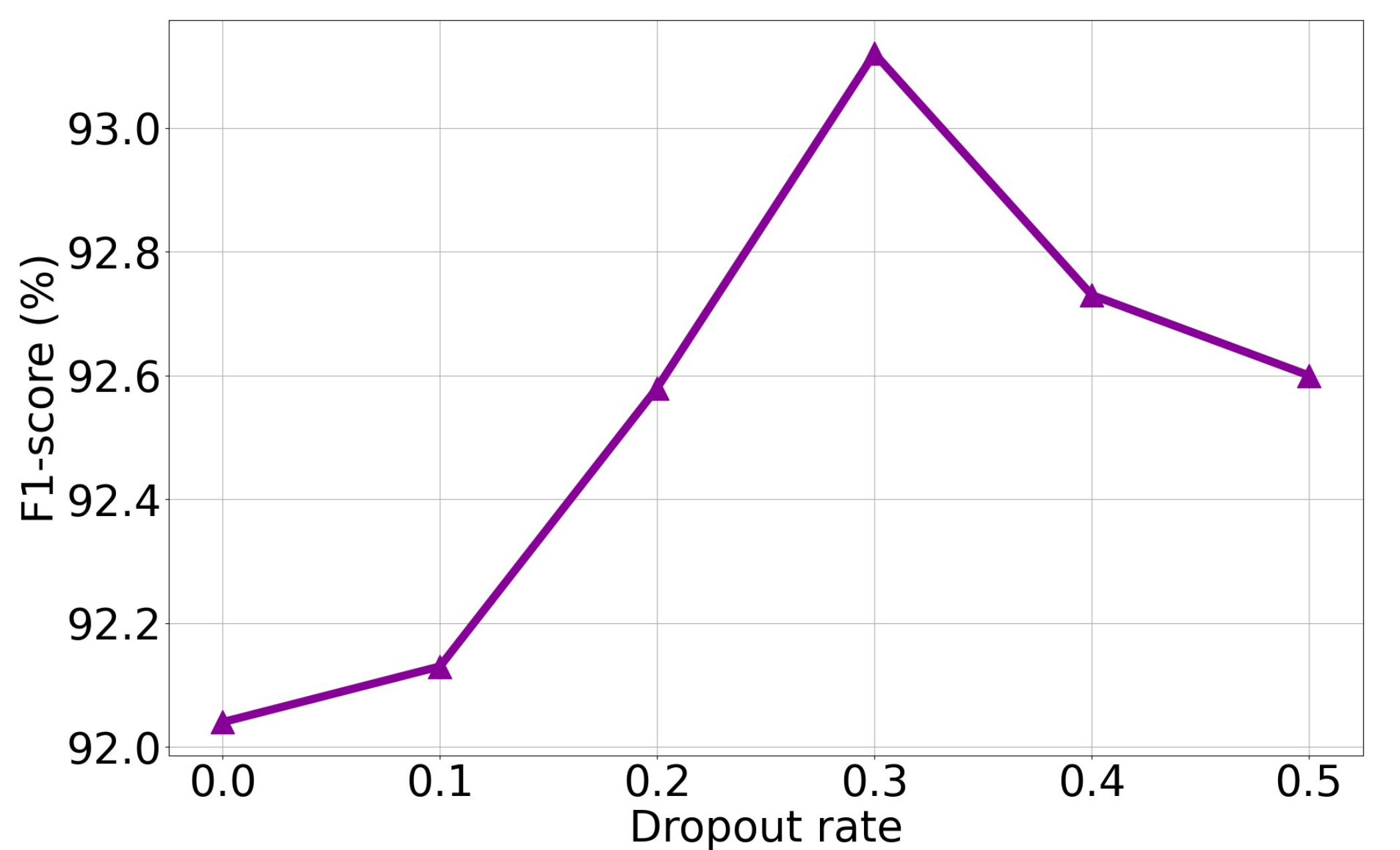}
    \caption{}
    \label{fig:subdrop}
  \end{subfigure}%
  \caption{F1-score with respect to the (a) $\alpha$ and $\gamma$ parameters in Focal Loss and (b) Dropout rate}
  \label{fig:DropBatch}
\end{figure*}

Another parameter that helps to improve the GRU architecture generalizability is the dropout rate. As illustrated in Fig.~\ref{fig:subdrop}, we investigated the effect of different dropout rates, ranging from 0 to 0.5, and identified 0.3 as the optimal rate. Dropout rates below this optimal threshold are associated with overfitting, where the model learns the noise in the training data too well, reducing its ability to generalize to unseen data. Conversely, dropout rates above 0.3 hinder effective learning, leading to an inadequate representation of the underlying distribution of the data. 

We further investigated how different $\alpha$ and $\gamma$ parameters in the Focal Loss function affected the model performance for imbalanced classification tasks on the validation set.
As demonstrated in Fig.~\ref{fig:subalpha}, varying levels of $\alpha$ and $\gamma$ show distinct impacts on the F1-score.
Results show that the optimal performance is achieved with $\alpha$ at 0.3 and $\gamma$ at 2, indicating a balanced approach to addressing class imbalance and focusing on hard examples. With $\gamma = 0$, where the loss is equivalent to Cross-Entropy, the importance given to the majority class ($\alpha$) is best at 0.2, which is close to the ratio of the minority (Normal) class in our dataset.
Interestingly, increasing $\gamma$ to 5 does not consistently improve model performance, contrary to expectations that higher $\gamma$ values would more aggressively penalize easy examples and thus benefit model learning from hard examples. 
This trend suggests a threshold beyond which increasing $\gamma$ may not yield additional benefits and could potentially lead to overfitting on hard examples or under-representing easy examples.
Furthermore, the performance improvement observed with increasing $\alpha$ values up to 0.3, followed by a slight decline, highlights the importance of $\alpha$ in modulating the loss contribution of different classes. It suggests an optimal balance at $\alpha = 0.3$, where the model sufficiently addresses class imbalance without disproportionately penalizing the majority class or overlooking the minority class.

To evaluate the performance of the proposed framework, we use the metrics as in Eq.~\ref{eq:metrics1} and~\ref{eq:metrics2}.

{
% \footnotesize 
% \scriptsize 
\begin{equation}
    \label{eq:metrics1}
    \begin{split}
        \mathrm{ACC} = &~\frac{TP+TN}{TP+TN+FP+FN}. \\ 
        \mathrm{SPE} = &~\frac{TN}{TN+FP}. \\
        \mathrm{SEN} = &~\frac{TP}{TP+FN}, \\
        \mathrm{PRC} = &~\frac{TP}{TP+FP}, \\
        \mathrm{F1-score} = &~\frac{2\times \mathrm{SEN}\times \mathrm{PRC}}{\mathrm{SEN}+\mathrm{PRC}}.\\
    \end{split}
\end{equation}
}

\begin{equation}
  \label{eq:metrics2}
  \resizebox{.9\columnwidth}{!}{$
      \mathrm{MCC} = \frac{TP \times TN - FP \times FN}{\sqrt{(TP+FP)(TP+FN)(TN+FP)(TN+FN)}},
  $}
\end{equation}
where $TP$, $TN$, $FP$, and $FN$ denote true positive, true negative, false positive, and false negative, respectively. We also report area under receiver operating characteristic curve (AUC).

\subsection{Experimental Results}

In this section, we present the experimental results of our proposed end-to-end framework for glaucoma classification using 3D OCT images. We compared our method against two approaches: a 3D Convolutional Neural Network (3D-CNN) specifically designed for 3D OCT images and the RETFound model encoder extended with two FC layers for individual 2D slice classification.

\begin{table*}[!htbp]
\centering
\caption{Summary of glaucoma detection results averaged from a 5-fold cross-validation with 95\% confidence intervals (±). Values are reported as percentages, with the confusion matrix relative to the total samples per class.}
\label{tab:results}

\begin{tabular}{lccccccccc}
\hline
\multicolumn{1}{c}{Method}  & ACC & AUC  & SEN  & SPE  & PRC  & F1-score  & MCC  & \multicolumn{2}{l}{Confusion Matrix} \\ \hline
3D-CNN~\cite{Maetschke2019feature} & 77.62 (± 9.78)          & 90.91 (± 3.34)          & 82.59                & 53.93                & 89.61                & 85.96                & \multicolumn{1}{c|}{32.41}          & 89.61      & 10.39      \\    & \multicolumn{1}{l}{}    & \multicolumn{1}{l}{}    & \multicolumn{1}{l}{} & \multicolumn{1}{l}{} & \multicolumn{1}{l}{} & \multicolumn{1}{l}{} & \multicolumn{1}{l|}{}               & 60.84         & 39.16      \\
RETFound~\cite{zhou2023foundation} & 83.51 (± 2.26)          & 88.35 (± 1.73)          & 87.95                & 67.01                & 90.84                & 89.37                & \multicolumn{1}{c|}{52.81}          & 90.85       & 9.15    \\    & \multicolumn{1}{l}{}    & \multicolumn{1}{l}{}    & \multicolumn{1}{l}{} & \multicolumn{1}{l}{} & \multicolumn{1}{l}{} & \multicolumn{1}{l}{} & \multicolumn{1}{l|}{}               & 40.11     & 59.89    \\
Proposed (ViT-large + GRU)                               & \textbf{89.19 (± 1.89)} & \textbf{94.20 (± 2.00)} & \textbf{91.83}       & \textbf{79.67}       & \textbf{94.21}       & \textbf{93.01}       & \multicolumn{1}{c|}{\textbf{69.33}} & 94.21      & 5.79       \\     & \multicolumn{1}{l}{}    & \multicolumn{1}{l}{}    & \multicolumn{1}{l}{} & \multicolumn{1}{l}{} & \multicolumn{1}{l}{} & \multicolumn{1}{l}{} & \multicolumn{1}{l|}{}               & 26.98       & 73.02      \\ \hline
\end{tabular}
\end{table*}

\begin{figure*}[!htbp]
    \centering
    \parbox[b]{0.14\textwidth}{\centering Slice 1} \hfill
    \parbox[b]{0.14\textwidth}{\centering Slice 32} \hfill
    \parbox[b]{0.14\textwidth}{\centering Slice 64} \hspace{2em}
    \parbox[b]{0.14\textwidth}{\centering Slice 1} \hfill
    \parbox[b]{0.14\textwidth}{\centering Slice 32} \hfill
    \parbox[b]{0.14\textwidth}{\centering Slice 64} \\

    \vspace{0.5em} % Space between descriptions and first row of images
    % First row: Subfigures 1-6
    %
    \includegraphics[width=0.14\textwidth]{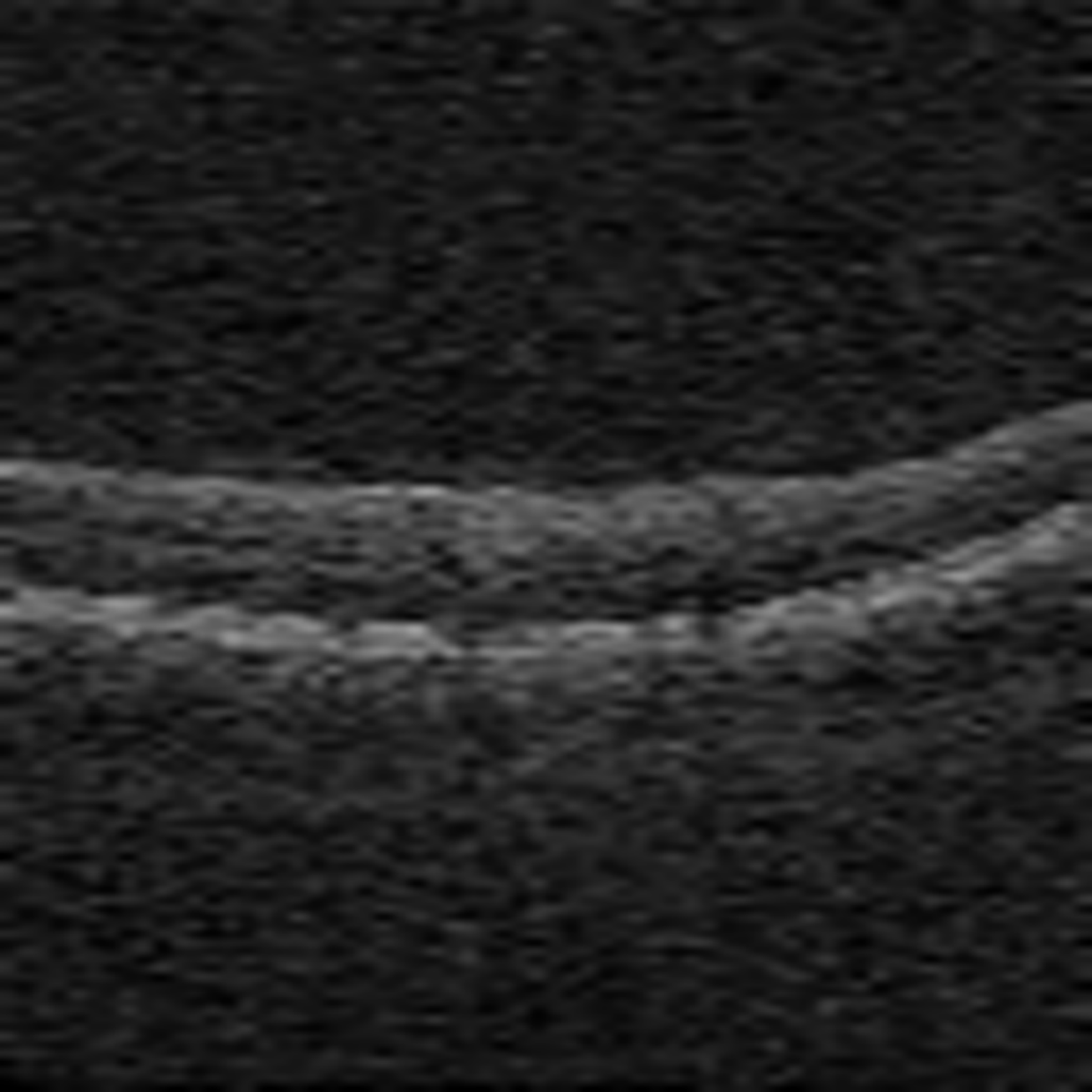} \hfill
    \includegraphics[width=0.14\textwidth]{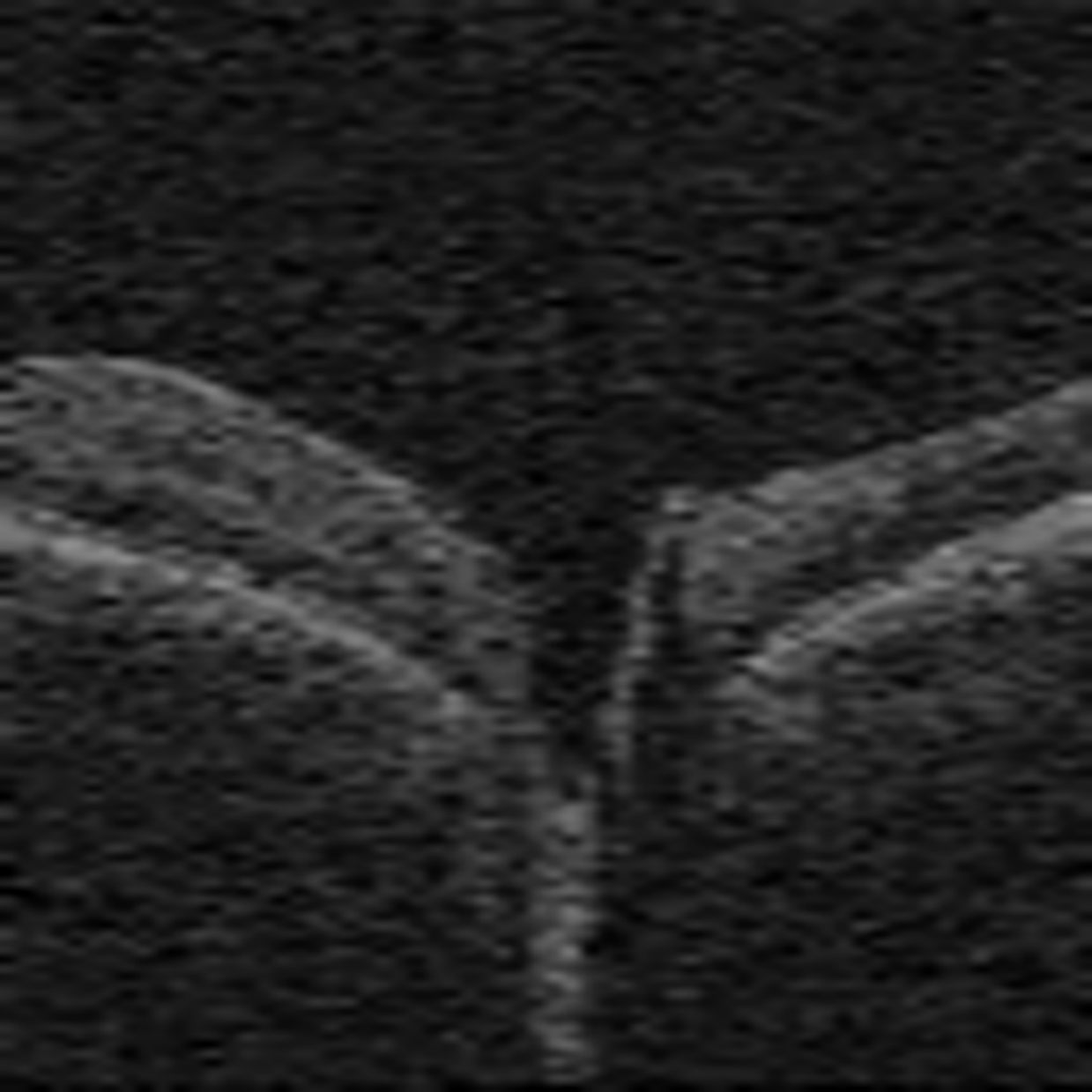} \hfill
    \includegraphics[width=0.14\textwidth]{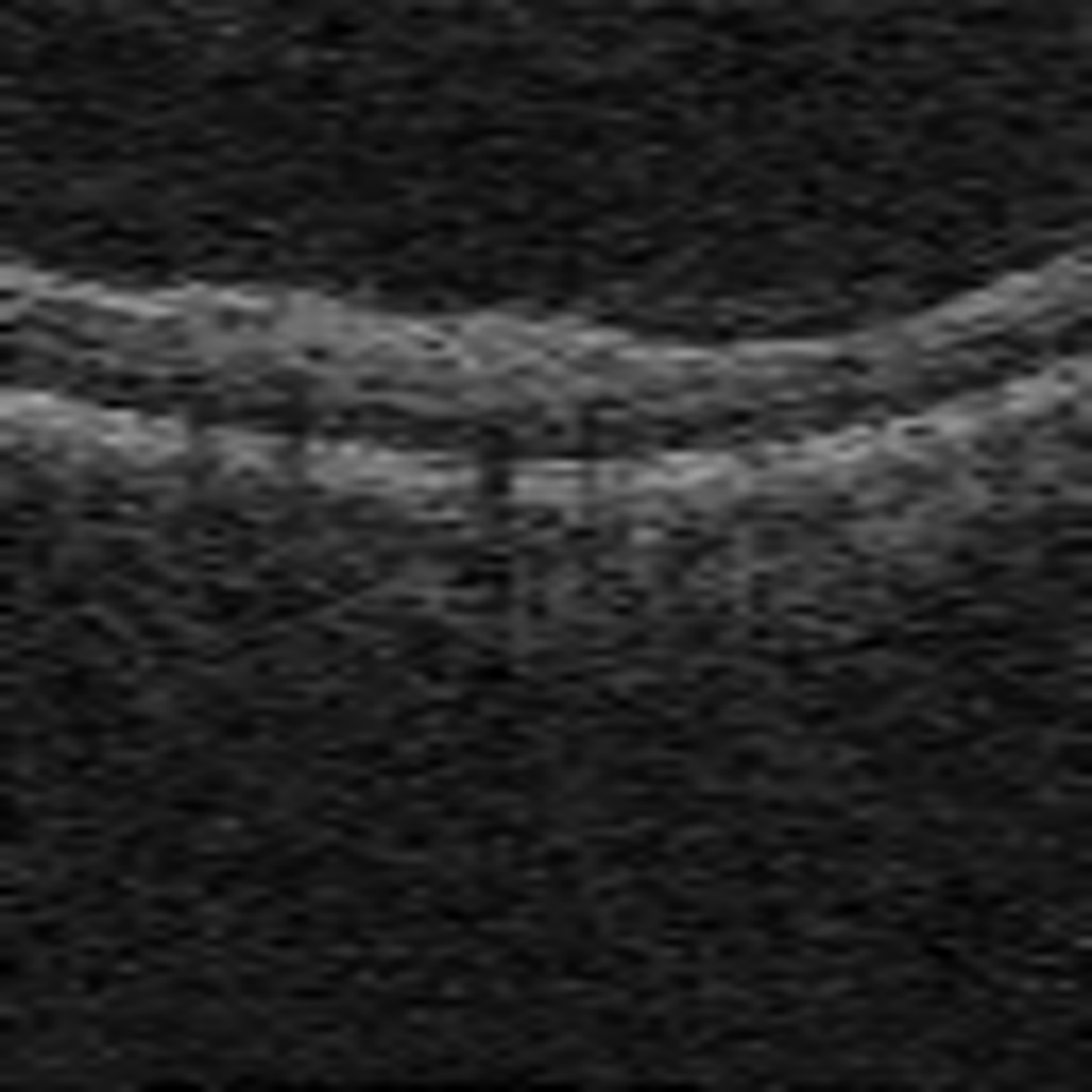} \hspace{2em}
    \includegraphics[width=0.14\textwidth]{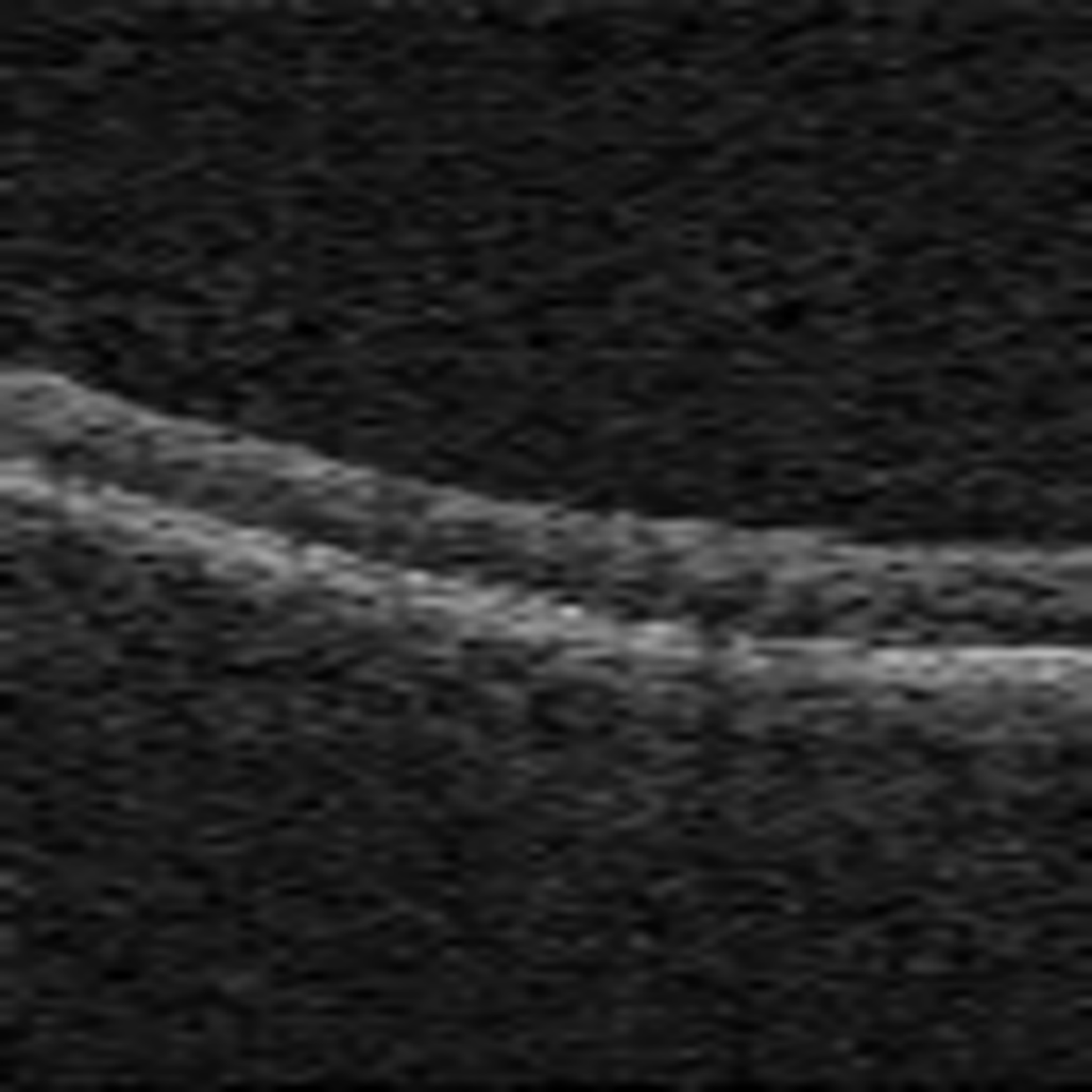} \hfill
    \includegraphics[width=0.14\textwidth]{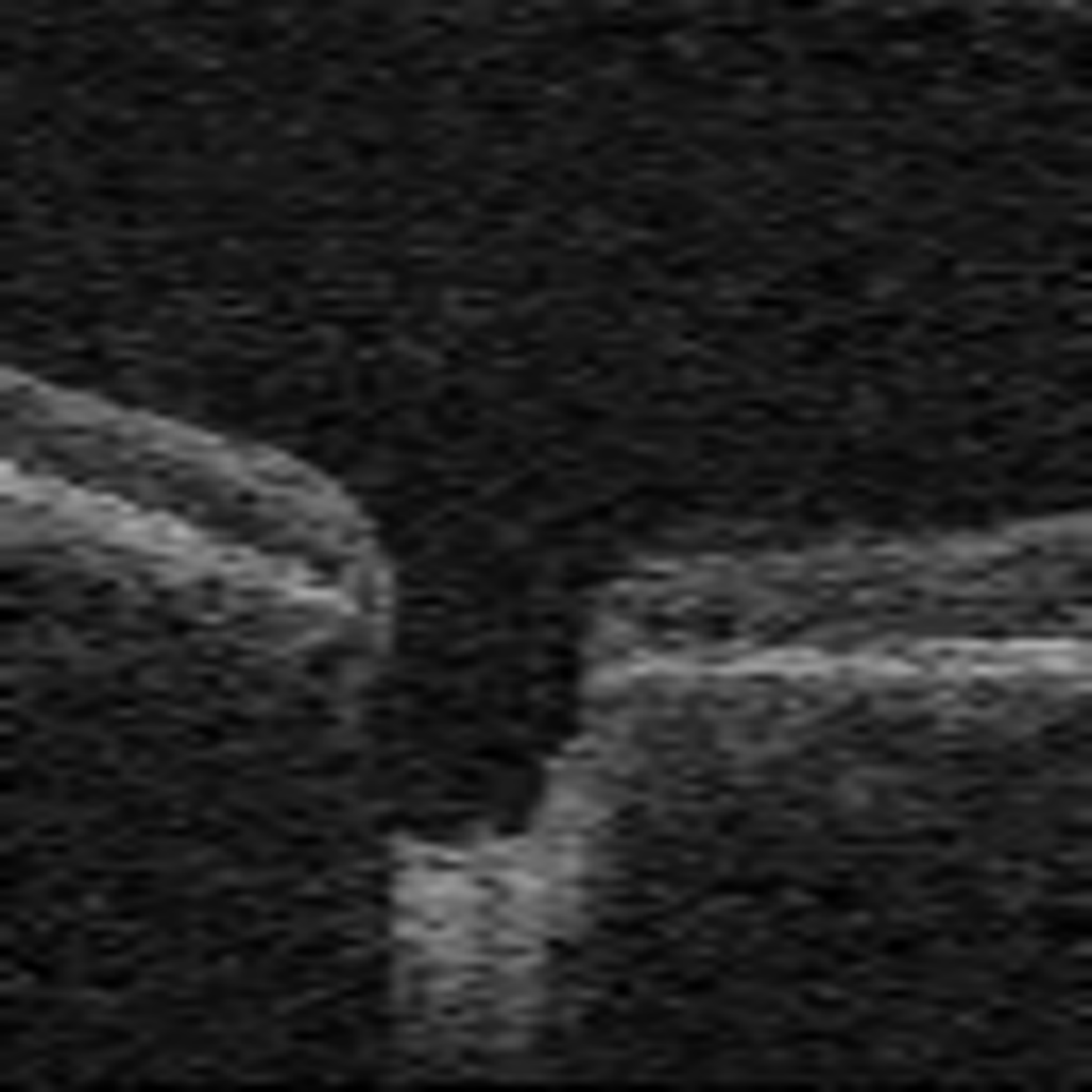} \hfill
    \includegraphics[width=0.14\textwidth]{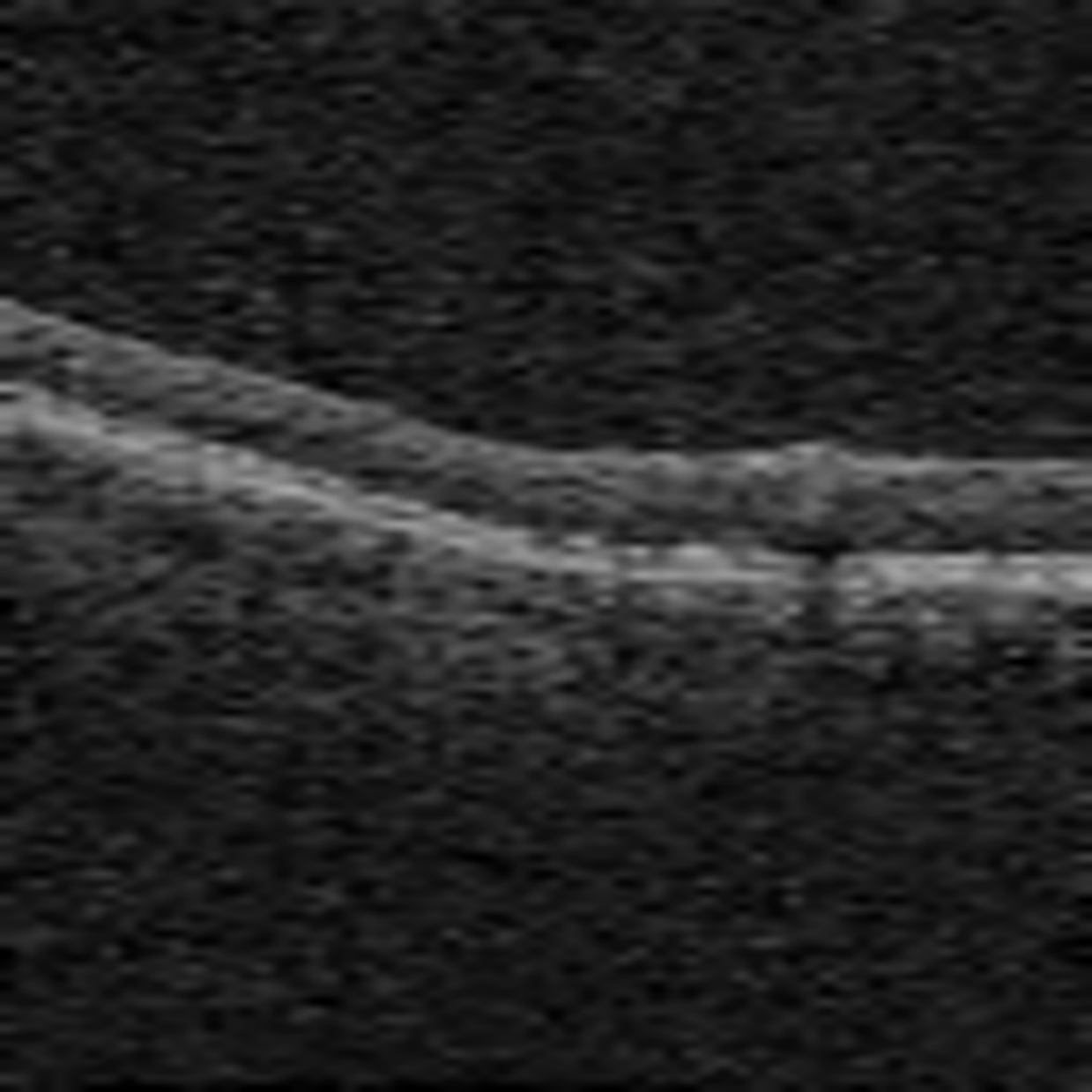} \\
    
    \vspace{1.2em} % Space between rows

    % Second row: Subfigures 7-12
    \includegraphics[width=0.14\textwidth]{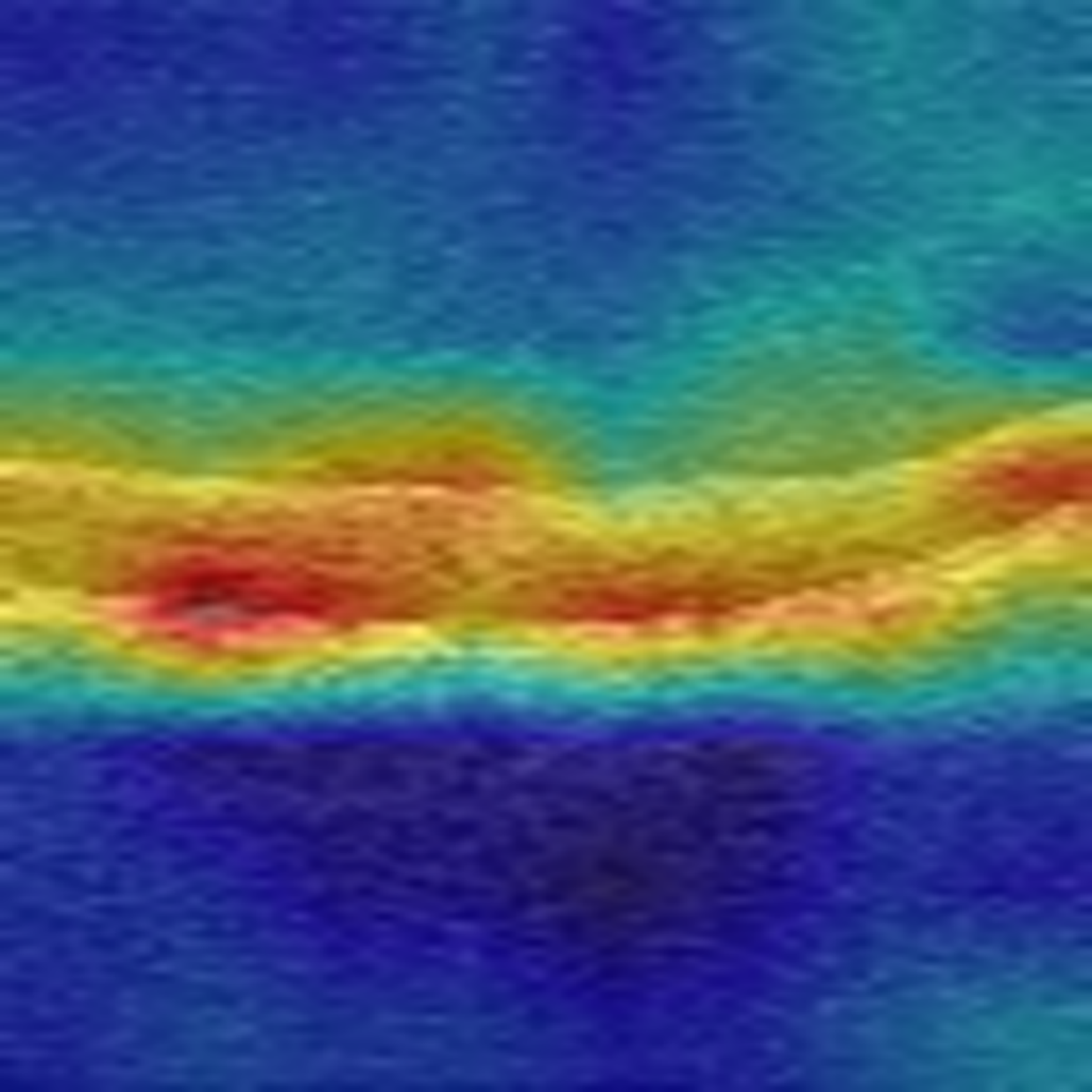}
    \hfill
    \subcaptionbox{Healthy}{%
    \includegraphics[width=0.14\textwidth]{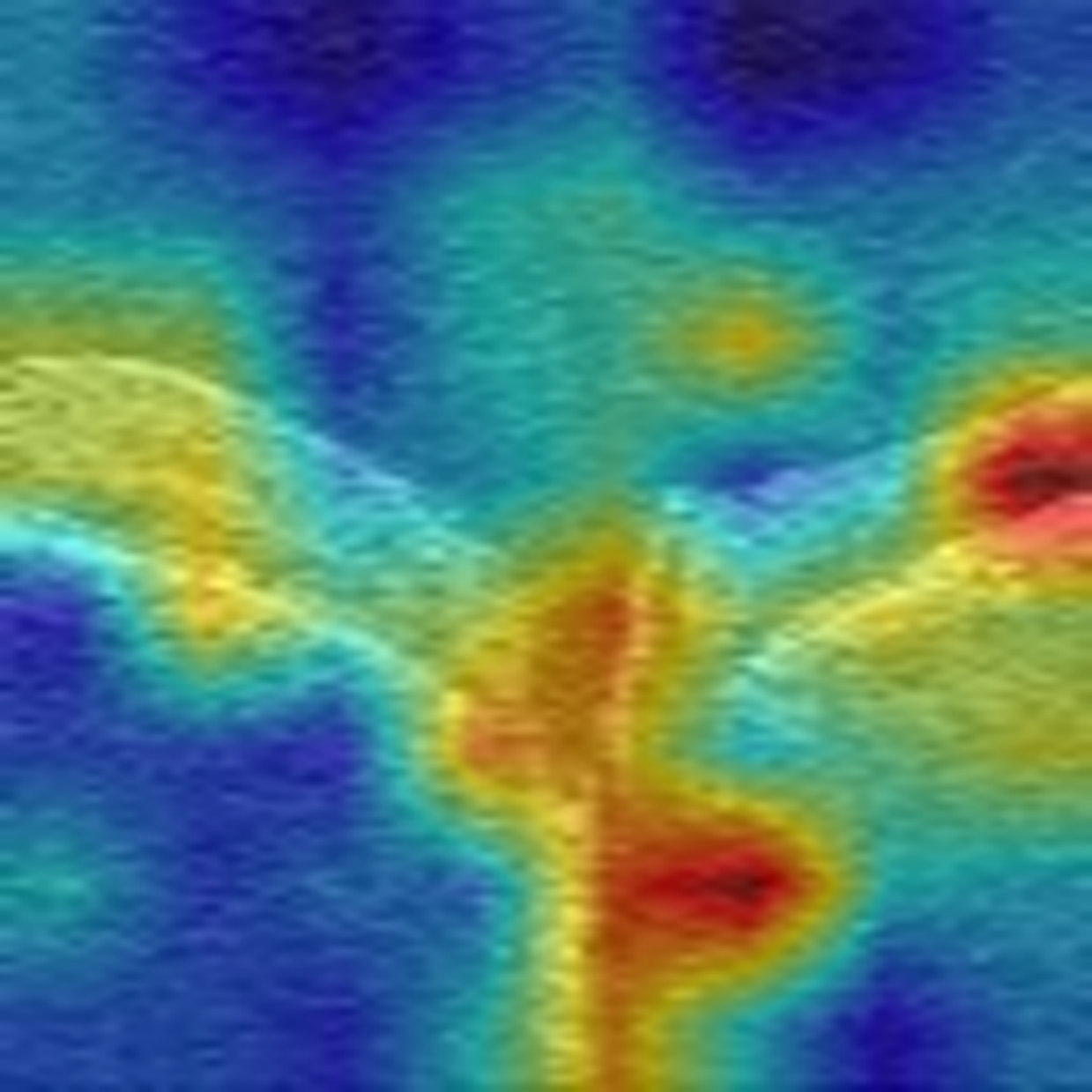}}
    \hfill
    \includegraphics[width=0.14\textwidth]{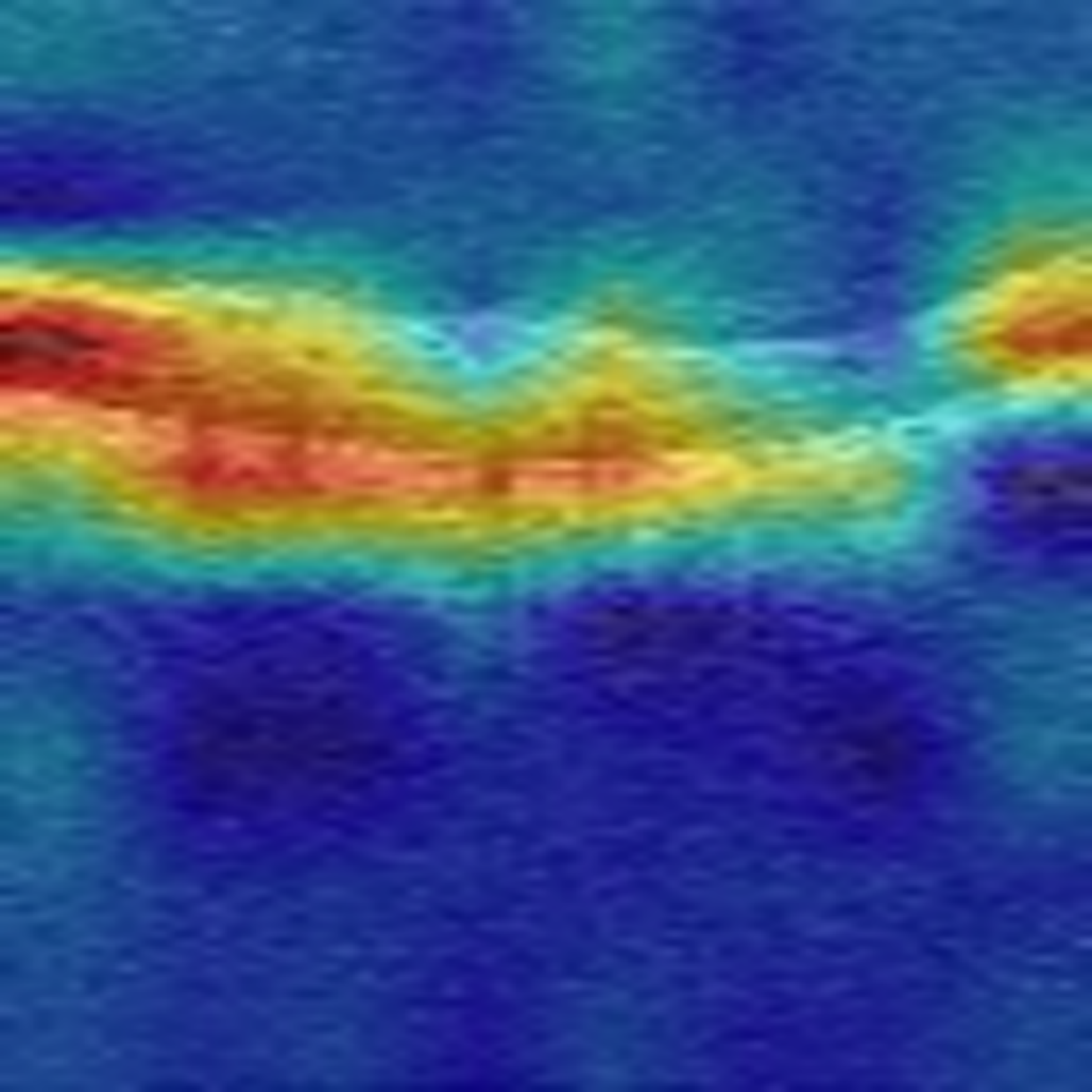}
    \hspace{2em}
    \includegraphics[width=0.14\textwidth]{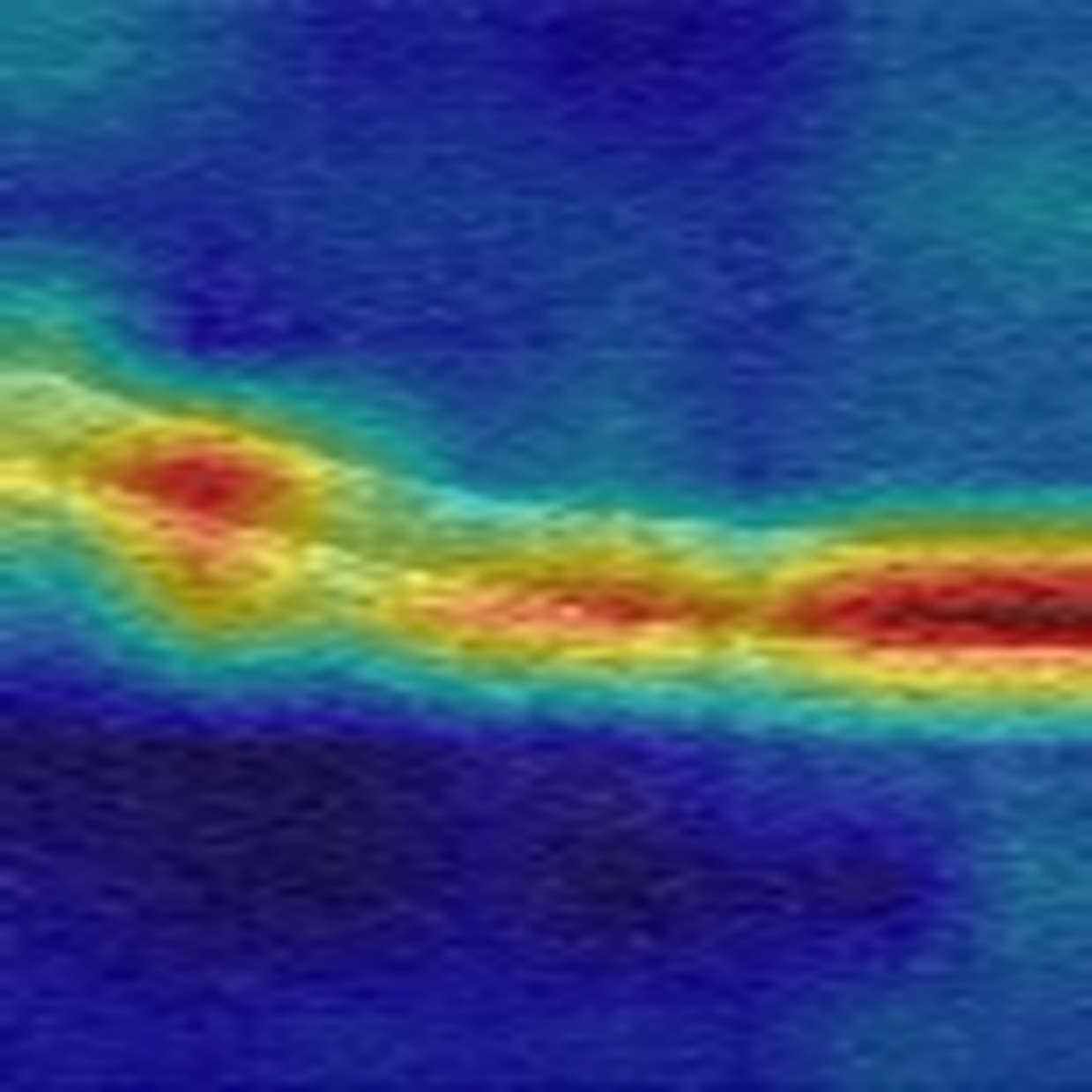}
    \hfill
    \subcaptionbox{Glaucoma}{%
    \includegraphics[width=0.14\textwidth]{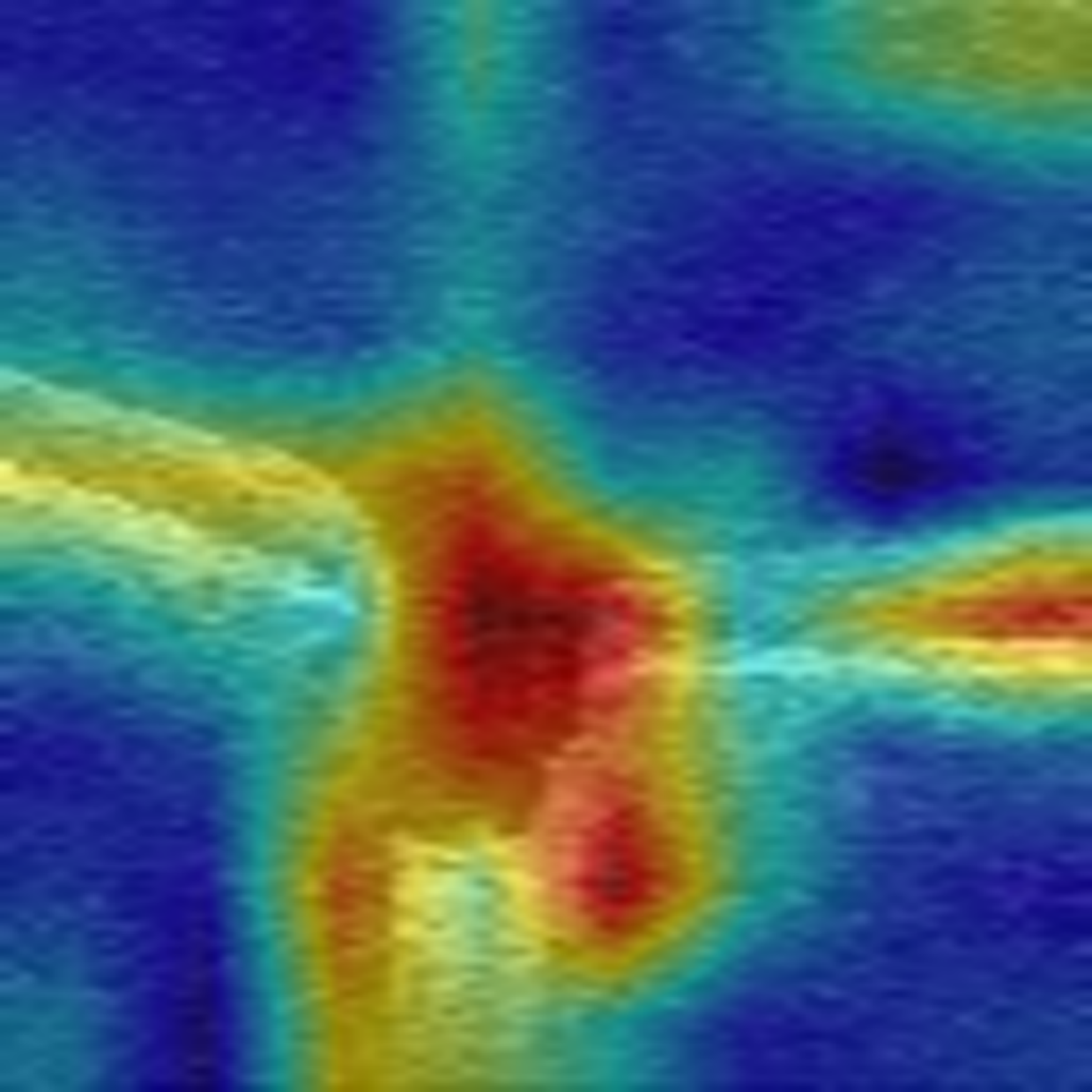}}
    \hfill
    \includegraphics[width=0.14\textwidth]{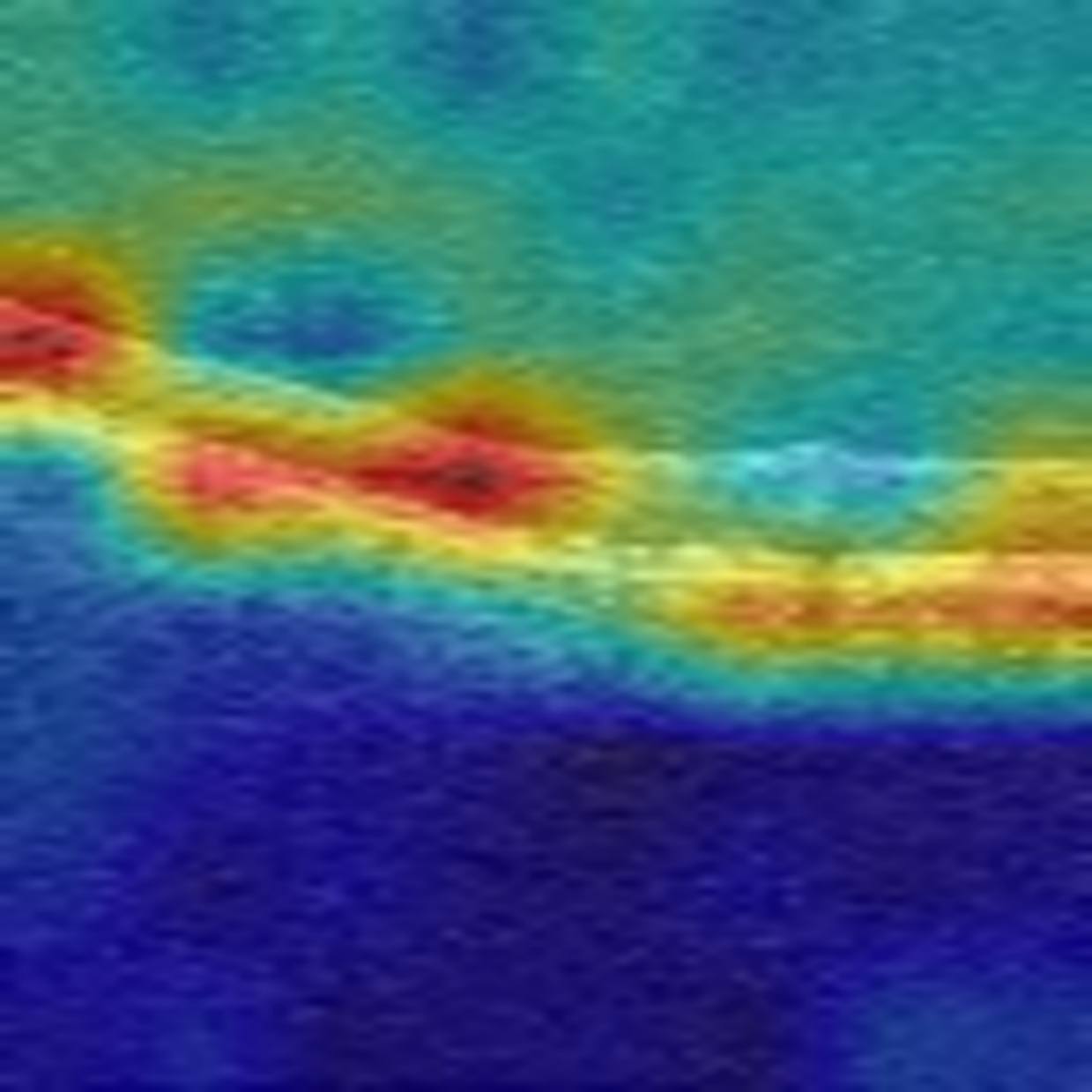}
    % \hfill
    
    \caption{Top row: Original 3D OCT image slices. Bottom row: Corresponding attention heatmaps generated using the attention rollout method~\cite{abnar2020Quantifying}. (a) Healthy scans (left 3 columns) and (b) Glaucoma scans (right 3 columns). Columns represent selected slices from the 3D OCT volume (Slice 1, Slice 32, Slice 64)}
    \label{fig:glaucoma_vs_healthy}
\end{figure*}

To evaluate the performance of our proposed framework, we first froze all the pre-trained layers within our model's backbone feature extractors that were trained on OCT images. 
This strategy was employed to leverage the rich, learned representations specific to the OCT image domain, thus ensuring the extraction of precise and relevant features for glaucoma classification. We then trained our proposed framework using a 5-fold cross-validation method. The performance metrics averaged across all folds are meticulously presented in Table~\ref{tab:results}, offering a comprehensive overview of the model's effectiveness in classifying glaucoma. Additionally, we used the attention rollout method~\cite{abnar2020Quantifying} to visualize the cumulative attention across all layers of the ViT-large backbone, highlighting informative regions in OCT images (Fig.~\ref{fig:glaucoma_vs_healthy}). The heatmaps reveal retina-specific anatomical structures, such as the optic nerve and retinal nerve fiber layer, in warmer colors, indicating their higher contribution to the extracted features.

The 3D-CNN architecture used for comparative analysis is identical to that proposed by Maetschke et al.~\cite{Maetschke2019feature}. Additionally, we extended the RETFound model encoder by integrating two FC layers with 256 and 64 nodes, respectively. This model was designed to process 2D slices from the 3D OCT scans, each annotated with subject-level labels, thereby aligning with the dimensional requirements of our study. To ensure a fair and rigorous comparison, both of these architectures underwent the same 5-fold cross-validation training approach as our proposed framework. This consistency ensured that the same instances were used across each fold, thereby facilitating an equitable and direct evaluation against our framework.

As summarized in Table~\ref{tab:results}, our proposed framework significantly outperforms the other methods across all evaluation metrics. Specifically, our model achieved an accuracy of 89.19\%, which is substantially higher than the 77.62\% accuracy of the 3D-CNN model and the 83.51\% accuracy of the standalone RETFound model. Moreover, the 3D-CNN model and standalone RETFound model exhibit wide accuracy confidence intervals of ±9.78\% and ±2.26\%, respectively, derived from the 5-fold cross-validation process. This indicates a lower level of reliability and variable performance across different data subsets, underscoring uncertainty in their predictive capabilities. The superior AUC of 94.20\% further emphasizes the proposed model's outstanding ability to distinguish between glaucomatous and non-glaucomatous scans.

Importantly, despite the challenges posed by an imbalanced dataset, the proposed model demonstrated a remarkable capability to accurately identify true positives and true negatives, achieving sensitivity and specificity scores of 91.83\% and 79.67\%, respectively. This performance is particularly noteworthy when compared to the 3D-CNN model, which showed a clear bias towards the majority class, as evidenced by its significantly lower specificity score of 53.93\%. The precision, F1-score, and Matthews Correlation Coefficient (MCC) further substantiate the robustness of the proposed model, with scores that notably outperform those of the 3D-CNN and standalone RETFound models.

These results underscore our model's nuanced capability to capture both the local and global structural integrity of the retina through the integration of slice-based feature extraction and sequential analysis, thereby ensuring a comprehensive understanding of the intricate patterns associated with glaucoma.

\subsection{Ablation Study}

We evaluated the contribution of individual components of the proposed framework in glaucoma detection. 
This experiment involves selectively replacing components and assessing the resulting impact on model effectiveness, thereby identifying the most crucial parts for achieving optimal outcomes.
Initially, we replaced the feature extraction component, which is based on the pre-trained ViT-large model, with a pre-trained ResNet34~\cite{He2016Residual} model. While our ViT-large backbone feature extractor is a specialized model pre-trained on OCT images, ResNet34 has been broadly pre-trained on the ImageNet dataset, leading to an inferior performance as depicted in Table~\ref{tab:ablation_results}. This outcome highlights the importance of task-specific model training.

\begin{table*}[!htbp]
\centering
\caption{Summary of ablation study results averaged from a 5-fold cross-validation with 95\% confidence intervals (±). Values are reported as percentages, with the confusion matrix relative to the total samples per class.}
\label{tab:ablation_results}
\begin{tabular}{lccccccccc}
\hline
\multicolumn{1}{c}{Method}        & ACC                  & AUC                  & SEN                  & SPE                  & PRC                  & F1-score             & MCC                        & \multicolumn{2}{l}{Confusion Matrix} \\ \hline
ResNet34 + GRU                    & 87.21 (± 2.22)       & 93.55 (± 2.53)       & 90.85                & 72.86                & 92.56                & 91.70                & \multicolumn{1}{c|}{63.92} & 92.56   & 7.44            \\    & \multicolumn{1}{l}{} & \multicolumn{1}{l}{} & \multicolumn{1}{l}{} & \multicolumn{1}{l}{} & \multicolumn{1}{l}{} & \multicolumn{1}{l}{} & \multicolumn{1}{l|}{}      & 30.04              & 69.96             \\
ViT-large + LSTM   & 87.51 (± 2.88)     & 93.63 (± 2.64) & 
91.65 &	73.85 &	91.97 &	91.81 & \multicolumn{1}{c|}{65.23}
& 91.97  & 8.03           \\    & \multicolumn{1}{l}{} & \multicolumn{1}{l}{} & \multicolumn{1}{l}{} & \multicolumn{1}{l}{} & \multicolumn{1}{l}{} & \multicolumn{1}{l}{} & \multicolumn{1}{l|}{}      & 27.00             & 73.00            \\
ViT-large + SVM (Slice 1)         & 77.32 (± 2.95)       & 83.75 (± 2.35)       & 91.38                & 51.41                & 77.57                & 83.91                & \multicolumn{1}{c|}{48.06} & 77.57          & 22.43               \\    & \multicolumn{1}{l}{} & \multicolumn{1}{l}{} & \multicolumn{1}{l}{} & \multicolumn{1}{l}{} & \multicolumn{1}{l}{} & \multicolumn{1}{l}{} & \multicolumn{1}{l|}{}      & 23.58             & 76.42           \\
ViT-large + SVM (Slice 31)        & 78.46 (± 4.46)       & 85.55 (± 4.03)       & 91.19                & 53.23                & 79.46                & 84.92                & \multicolumn{1}{c|}{49.31} & 79.47    & 20.53            \\     & \multicolumn{1}{l}{} & \multicolumn{1}{l}{} & \multicolumn{1}{l}{} & \multicolumn{1}{l}{} & \multicolumn{1}{l}{} & \multicolumn{1}{l}{} & \multicolumn{1}{l|}{}      & 24.71                &  75.29            \\
ViT-large + SVM (Slice 32)        & 79.42 (± 3.86)       & 87.30 (± 2.83)       & 91.54                & 54.79                & 80.52                & 85.68                & \multicolumn{1}{c|}{51.20} & 80.53   & 19.47               \\    & \multicolumn{1}{l}{} & \multicolumn{1}{l}{} & \multicolumn{1}{l}{} & \multicolumn{1}{l}{} & \multicolumn{1}{l}{} & \multicolumn{1}{l}{} & \multicolumn{1}{l|}{}      & 23.95              & 76.05               \\
ViT-large + SVM (Slice 37)        & 79.10 (± 1.41)       & 86.58 (± 3.40)       & 91.61                & 54.18                & 79.93                & 85.37                & \multicolumn{1}{c|}{50.80} & 79.93    & 20.07               \\    & \multicolumn{1}{l}{} & \multicolumn{1}{l}{} & \multicolumn{1}{l}{} & \multicolumn{1}{l}{} & \multicolumn{1}{l}{} & \multicolumn{1}{l}{} & \multicolumn{1}{l|}{}      & 23.58              & 76.42             \\
ViT-large + SVM (Slice 52)        & 76.77 (± 2.77)       & 85.37 (± 2.88)       & 90.62                & 50.65                & 77.57                & 83.59                & \multicolumn{1}{c|}{46.20} & 77.57   & 22.43               \\    & \multicolumn{1}{l}{} & \multicolumn{1}{l}{} & \multicolumn{1}{l}{} & \multicolumn{1}{l}{} & \multicolumn{1}{l}{} & \multicolumn{1}{l}{} & \multicolumn{1}{l|}{}      & 25.86              & 74.14               \\
ViT-large + SVM (Majority voting) & 83.14 (± 1.87)       & 91.55 (± 2.10)       & 93.19                & 60.98                & 84.06                & 88.39                & \multicolumn{1}{c|}{59.02} & 84.05    & 15.95               \\     & \multicolumn{1}{l}{} & \multicolumn{1}{l}{} & \multicolumn{1}{l}{} & \multicolumn{1}{l}{} & \multicolumn{1}{l}{} & \multicolumn{1}{l}{} & \multicolumn{1}{l|}{}      & 19.77              & 80.23            \\ \hline
Proposed (ViT-large + GRU)        & 89.19 (± 1.89)       & 94.20 (± 2.00)       & 91.83                & 79.67                & 94.21                & 93.01                & \multicolumn{1}{c|}{69.33} & 94.21    & 5.79              \\    & \multicolumn{1}{l}{} & \multicolumn{1}{l}{} & \multicolumn{1}{l}{} & \multicolumn{1}{l}{} & \multicolumn{1}{l}{} & \multicolumn{1}{l}{} & \multicolumn{1}{l|}{}      & 26.98             & 73.02             \\ \hline
\end{tabular}
\end{table*}

\begin{figure*}[!htbp]
    \centering

    % Row 1
    \begin{subfigure}{0.32\textwidth}
        \includegraphics[width=\linewidth]{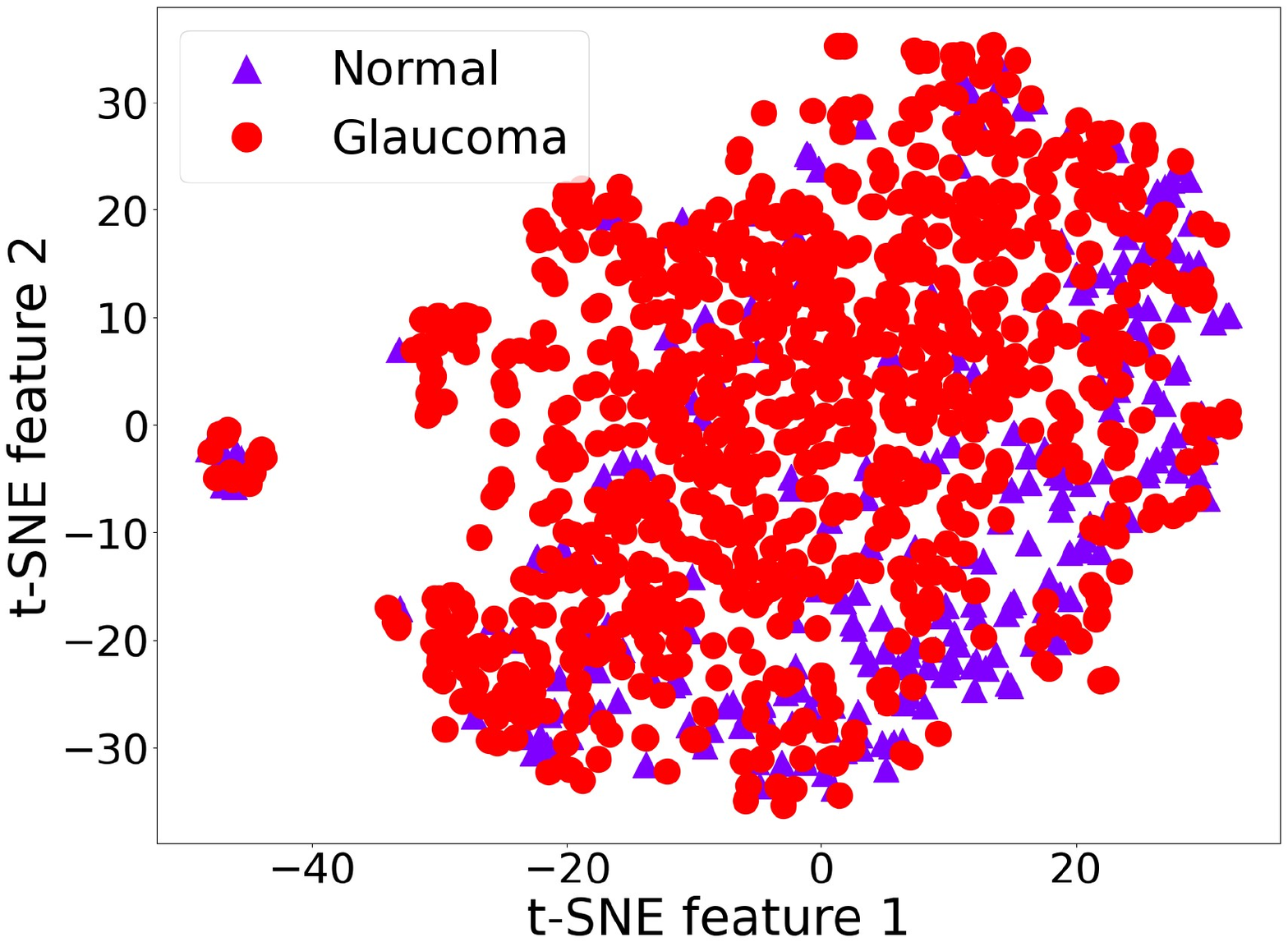}
        \caption{}
        \label{fig:1}
    \end{subfigure}
    \hfill
    \begin{subfigure}{0.32\textwidth}
        \includegraphics[width=\linewidth]{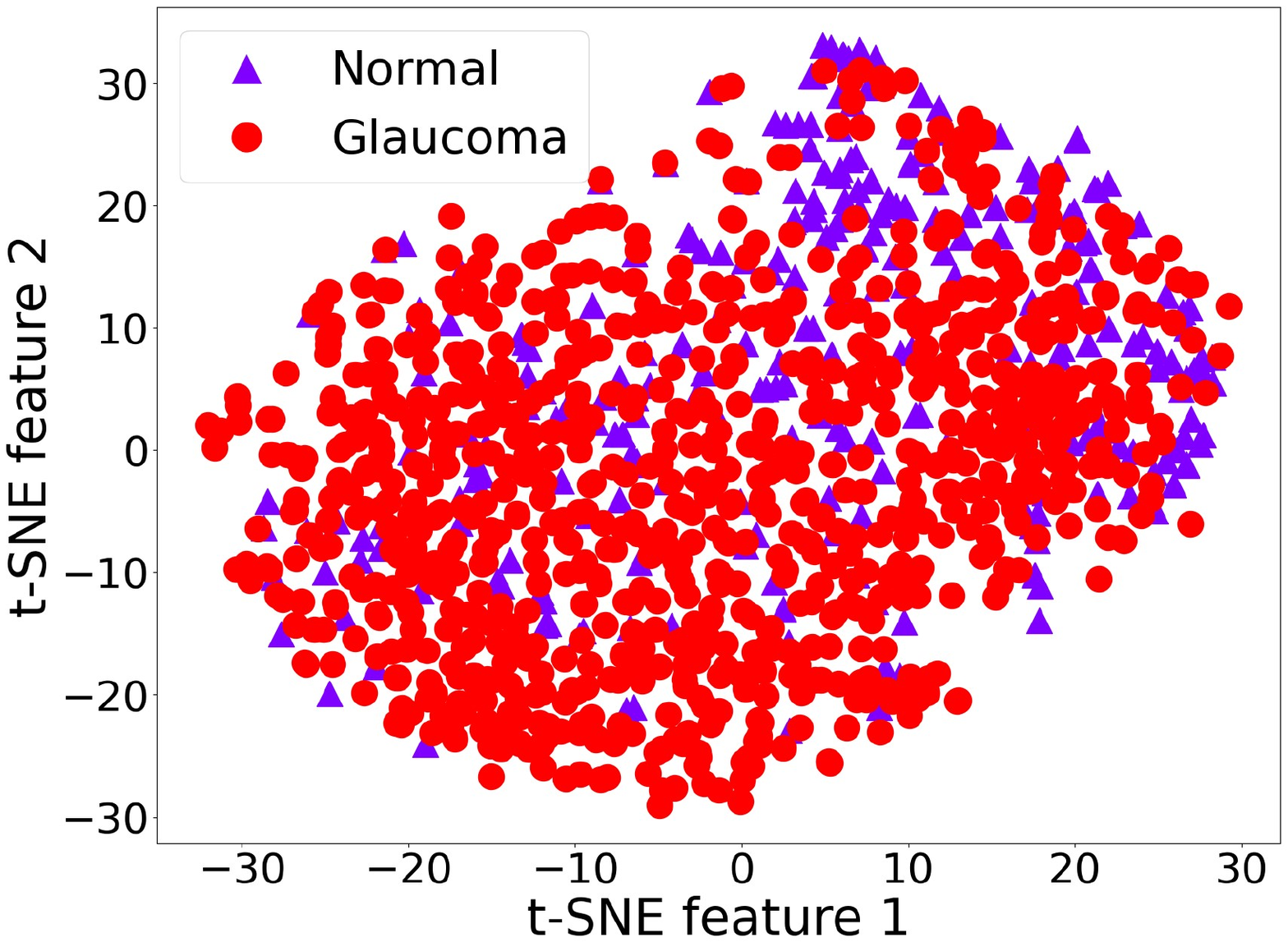}
        \caption{}
        \label{fig:2}
    \end{subfigure}
    \hfill
    \begin{subfigure}{0.32\textwidth}
        \includegraphics[width=\linewidth]{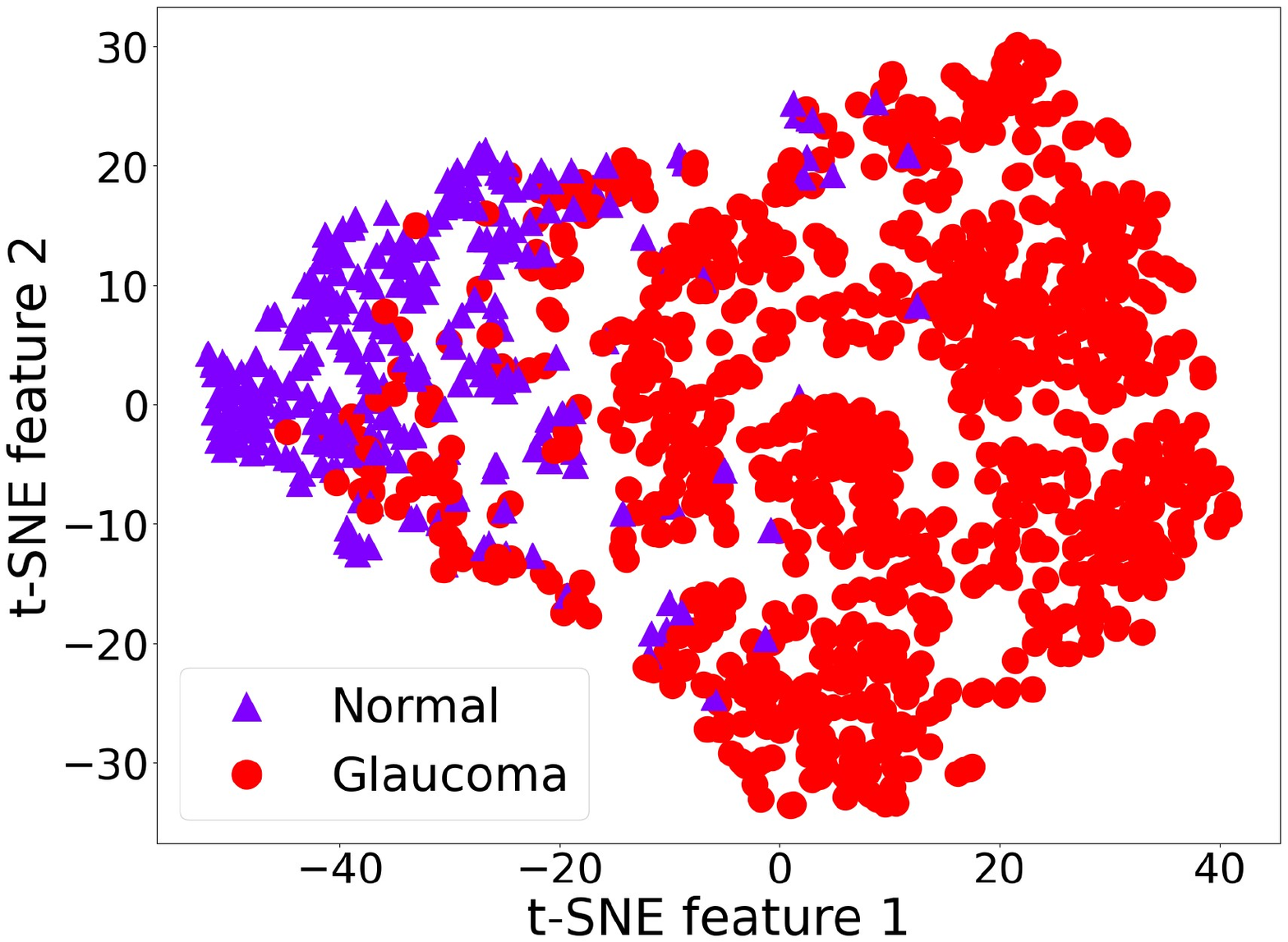}
        \caption{}
        \label{fig:3}
    \end{subfigure}

    \caption{t-SNE visualization of embedding spaces for (a) ViT-large output features for slice 32, (b) ResNet34 output features for slice 32, (c) ViT-large + GRU output features}
    \label{fig:tsne}
\end{figure*}

To further investigate the effect of different RNN layers on sequential dependency modeling, we replaced the GRU layers with LSTM layers. The LSTM employs a more complex gating mechanism to capture long-term dependencies more effectively. However, as reported in Table~\ref{tab:ablation_results}, the GRU layers marginally outperformed the LSTM layers. This result suggests that the simpler gating mechanism of the GRU reduces model complexity, making it more suitable for our framework.
In the subsequent experiment, we assessed the impact of replacing the sequential processing in the proposed framework with a voting ensemble model. To address the challenge of incorporating all available slices in 3D OCT images, we initially reduced the number of slices. Therefore, from the original set of 64 OCT slices, we chose 5 slices based on their entropy values. Entropy, in this context, measures the information content for identifying the most relevant slices for glaucoma detection. We computed the entropy of all slices in the OCT images. Considering the significance of the central slice in glaucoma diagnosis~\cite{WANG2020101695}, we selected four slices with entropy values close to the $32^{\mathrm{nd}}$ slice and the central slice itself. This method ensures a balance between data-driven selection and clinical relevance.

After extracting features from the five slices, we proceeded with feature selection to reduce the feature vector's dimensionality. We used the gain ratio method to reduce each slice's feature dimensions from 1024 to 128 individually. Each slice, now represented by the reduced feature set, was independently processed through a Support Vector Machine (SVM) to classify it. Additionally, we employed a majority voting ensemble technique to integrate the insights obtained from all five slices into a coherent diagnostic evaluation. The results of these experiments are detailed in Table~\ref{tab:ablation_results}.

We used t-distributed Stochastic Neighbor Embedding (t-SNE)~\cite{Laurens2008Visualizing} to explore the model outputs further. Specifically, for slice 32, we used t-SNE to visualize the output feature spaces generated by pre-trained ViT-large and ResNet34 feature extractors, as shown in Fig.~\ref{fig:1} and Fig.~\ref{fig:2}, respectively. By comparing the spatial distributions of features for the normal and glaucoma classes, we observed that the normal class samples were more densely clustered within the ViT-large feature space than in the ResNet34 feature space. This finding suggests that ViT-large has superior discriminative capabilities and is more effective in identifying and distinguishing the intricate patterns characteristic of glaucoma in OCT images.

We have visualized the feature space in our proposed framework, which is the last layer before the FC layer. This is shown in Fig.~\ref{fig:3}. The visualization clearly differentiates between the normal and glaucoma classes, which is a testament to the effectiveness of our framework. Our framework is highly capable of extracting clinically relevant features from OCT image slices and distinguishing sequential patterns to improve glaucoma detection accuracy.

\section{Conclusion}
\label{sec:conclusion}
In this study, we introduced a novel deep learning framework that addresses the critical challenge of accurate and early glaucoma detection from 3D OCT imaging. By leveraging the complementary strengths of a pre-trained Vision Transformer and a bidirectional GRU, the proposed approach enables rich features to be extracted from individual B-scan slices and capture inter-slice spatial dependencies associated with the ocular structure. The holistic analysis of local retinal characteristics and global structural integrity captures the intricate patterns associated with glaucoma.

The experimental results validate the framework's superior performance, achieving an AUC of 94.20\%, F1-score of 93.01\%, and MCC of 69.33\%. These metrics outperform state-of-the-art methods, including the 3D-CNN~\cite{Maetschke2019feature} and RETFound model~\cite{zhou2023foundation} extended for slice-based classification. These results underscore the potential of the proposed approach to address challenges posed by class imbalance and subtle glaucomatous patterns distributed across the ocular anatomy, enhancing clinical decision support systems and improving patient outcomes in glaucoma management.

In our opinion, there are several opportunities for future research. Exploring the integration of complementary clinical data modalities, such as visual field tests or patient demographics, could further enhance the diagnostic capabilities of the proposed framework. Moreover, investigating alternative sequence processing models and incorporating attention mechanisms may yield further performance improvements in glaucoma detection and screening tasks. Additionally, extending the application scope of our framework coupled with interpretability to other ocular pathologies could significantly contribute to the advancement of early diagnosis and treatment strategies within the broader field of ophthalmology. Furthermore, developing robust techniques to mitigate the impact of data imbalance could substantially enhance the diagnostic accuracy and generalizability of the framework across diverse datasets and imaging modalities. However, it is important to note that deploying this framework in clinical settings would require rigorous clinical trials, expert validation, and integration into real-world workflows that involve patients from diverse ethnicities, genders, and other demographic groups. Such strategies would enable a comprehensive assessment of the framework's robustness and facilitate its broader deployment in clinical ophthalmological practice.

\section*{References}
\vspace*{-1.2\baselineskip}
\bibliographystyle{IEEEtran}
\bibliography{references.bib}

\vspace*{-2\baselineskip}
\begin{IEEEbiography}[{\includegraphics[width=1in,height=1.25in,clip,keepaspectratio]{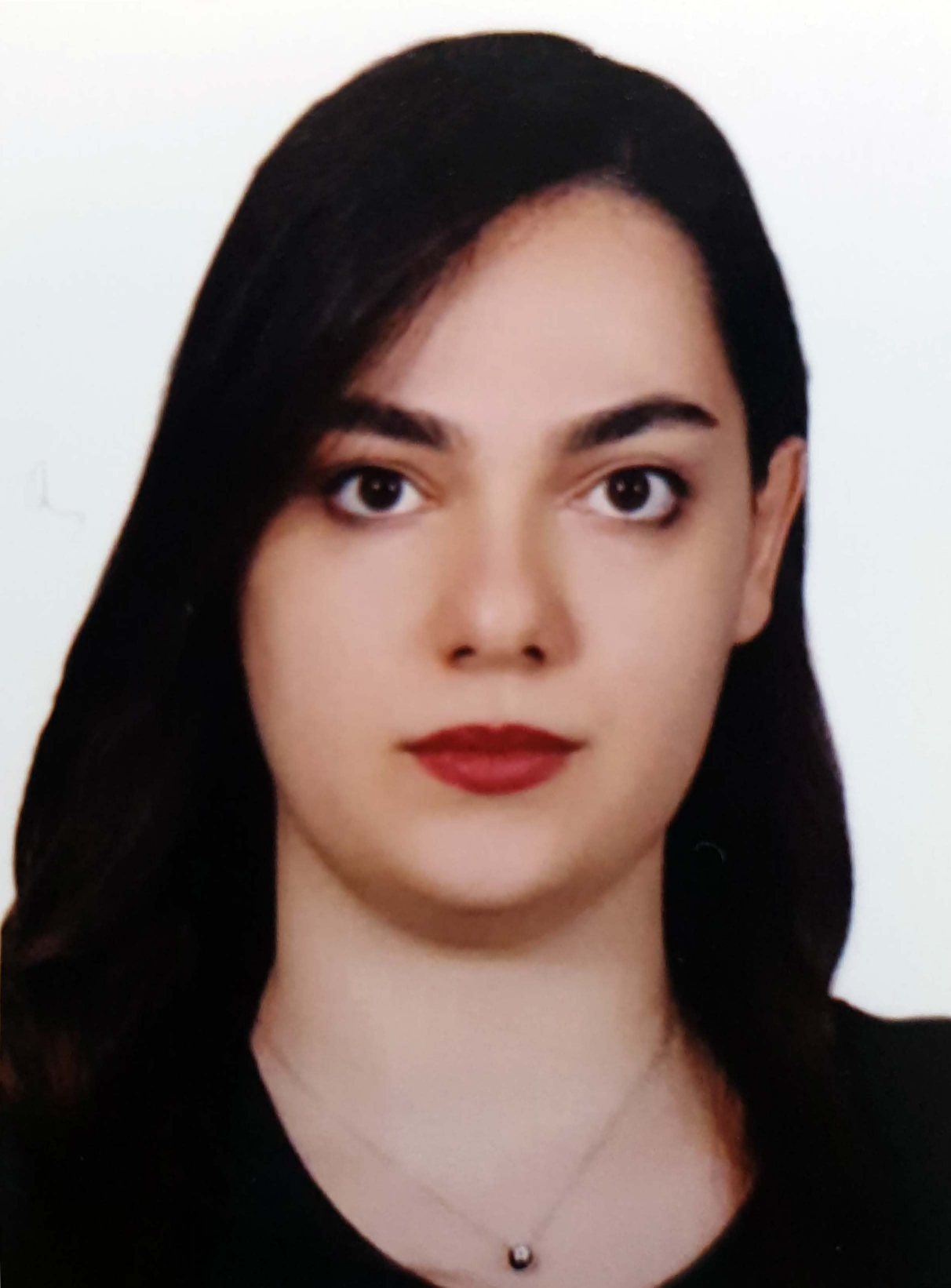}}]{Mona Ashtari-Majlan} completed her Master's degree in Health Systems Engineering at Amirkabir University of Technology in Tehran in 2021. She is currently a Ph.D. candidate in the Network and Information Technologies doctoral program at Universitat Oberta de Catalunya, Spain. Her research interests revolve around Computer Vision, Biomedical Image Processing, and Deep Learning.
\end{IEEEbiography}

\vspace*{-2\baselineskip}

\begin{IEEEbiography}[{\includegraphics[width=1in,height=1.25in,clip,keepaspectratio]{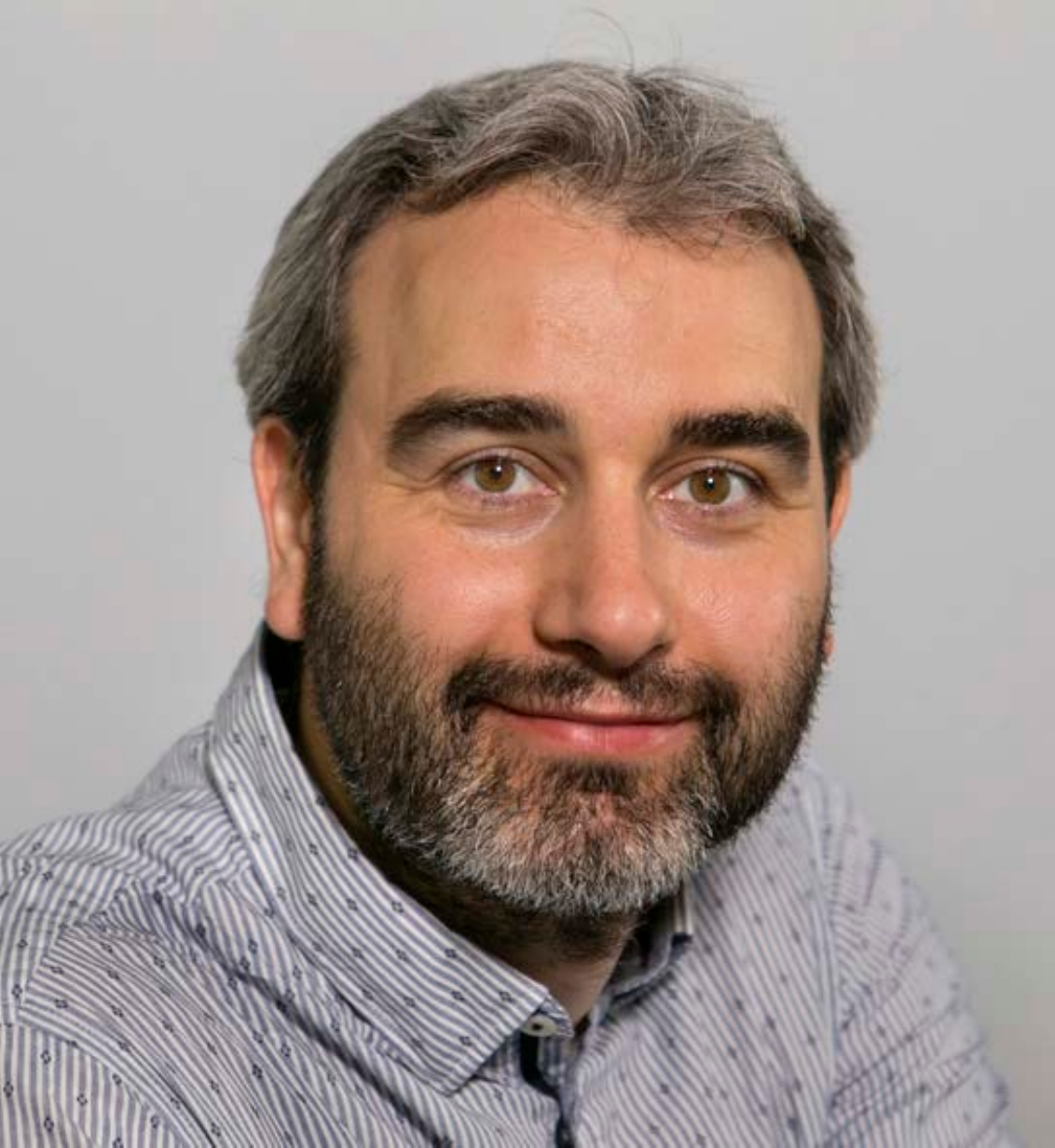}}]{David Masip} (Senior Member, IEEE) received his Ph.D. degree in Computer Vision in 2005 (Universitat Autonoma de Barcelona, Spain). He was awarded for the best thesis in Computer Science. He is a Full Professor at the Computer Science Multimedia and Telecommunications Department at Universitat Oberta de Catalunya, Spain, and the Director of the Doctoral School since 2015. He has published more than 70 scientific papers in relevant journals and conferences. His research interests include Oculomics, Retina Image Analysis, and Affective Computing.
\end{IEEEbiography}
\vfill

\end{document}